\documentclass[useAMS,usenatbib]{mn2e}

\usepackage{epsfig,graphicx,graphics}

\title[Cm-wave radiation from the $\rho$~Oph molecular
  cloud]{Centimeter-wave continuum radiation from the $\rho$~Ophiuchi molecular
  cloud} \author[Casassus et al.]{ Simon  Casassus$^{1}$\thanks{E-mail:
    simon@das.uchile.cl (SC)}, 
  Clive Dickinson$^{2}$, 
  Kieran Cleary$^{3}$, 
  Roberta Paladini$^{2}$,  \newauthor
  Mireya Etxaluze$^{4,5}$, 
  Tanya  Lim$^5$,
  Glenn J. White$^{4,5}$, 
  Michael Burton$^6$,  \newauthor  
  Balt  Indermuehle$^6$, 
  Otmar Stahl$^7$,
  Patrick  Roche$^8$  \\ 
  $^{1}$  Departamento de Astronom\'{\i}a, Universidad de Chile, Casilla 36-D,
  Santiago, Chile\\ 
  $^{2}$ Infrared Processing and Analysis Center, California Institute of Technology, M/S 220-6,  
1200 E. California Blvd., Pasadena, CA 91125. \\
  $^{3}$ Chajnantor Observatory, M/S 105-24,  California Institute of Technology, Pasadena, CA 91125 \\
  $^{4}$ Department of Physics and Astronomy, The Open University, Milton Keynes MK7 6AA, UK\\
  $^{5}$ The Rutherford Appleton Laboratory, Didcot, Oxfordshire OX11 0QX, UK. \\
  $^{6}$ School of Physics, University of New South Wales, Sydney NSW 2052, Australia\\
  $^{7}$ ZAH, Landessternwarte K\"onigstuhl, 69117 Heidelberg, Germany\\
  $^{8}$  Astrophysics, Oxford University, DWB, Keble Road, Oxford OX1
  3RH, UK\\
 }

\begin{document}

\date{}
\pagerange{\pageref{firstpage}--\pageref{lastpage}} \pubyear{2006}

\maketitle

\label{firstpage}

\begin{abstract}
The $\rho$~Oph molecular cloud is undergoing intermediate-mass star
formation. UV radiation from its hottest young stars heats and
dissociates exposed layers, but does not ionize hydrogen. Only faint
radiation from the Rayleigh-Jeans tail of $\sim$~10--100~K dust is
expected at wavelengths longwards of $\sim$3~mm. Yet Cosmic Background
Imager (CBI) observations reveal that the $\rho$~Oph~W
photo-dissociation region (PDR) is surprisingly bright at centimetre
wavelengths. We searched for interpretations consistent with the {\em
  WMAP} radio spectrum, new {\em ISO}-LWS parallel mode images and
archival {\em Spitzer} data.  Dust-related emission mechanisms at
1~cm, as proposed by Draine \& Lazarian, are a possibility. But a
magnetic enhancement of the grain opacity at 1~cm is inconsistent with
the morphology of the dust column maps $N_d$ and the lack of detected
polarization. Spinning dust, or electric-dipole radiation from
spinning very small grains (VSGs), comfortably explains the radio
spectrum, although not the conspicuous absence from the CBI data of
the infrared circumstellar nebulae around the B-type stars S~1 and
SR~3. Allowing for VSG depletion can marginally reconcile spinning
dust with the data. As an alternative interpretation we consider the
continuum from residual charges in $\rho$~Oph~W, where most of carbon
should be photoionised by the close binary HD147889 (B2IV,
B3IV). Electron densities of $\sim 10^2$~cm$^{-3}$, or H-nucleus
densities $n_\mathrm{H}>$~10$^6$~cm$^{-3}$, are required to interpret
$\rho$~Oph~W as the C\,{\sc ii} Str\"omgren sphere of
HD147889. However the observed steep and positive low-frequency
spectral index would then require optically thick emission from an
hitherto unobserved ensemble of dense clumps or sheets with a filling
factor $\sim 10^{-4}$ and $n_\mathrm{H} \sim 10^7$~cm$^{-3}$.

\end{abstract}

%The main source of
%residual charge in $\rho$~Oph~W is the photoionization of carbon by UV
%photons above 11.3~eV.


\section{Introduction}

The subtraction of Galactic foregrounds in experiments designed to map
the cosmic microwave background requires the examination of the
emission mechanisms at work in the interstellar medium (ISM). An
anomalous component of continuum emission was discovered in the
direction of Galactic cirrus clouds \citep{lei97}, so defined by their
H\,{\sc i}~21~cm and far-IR contours \citep[e.g.][]{bou88}.  The
LDN~1622 dark cloud \citep[Lynds Dark Nebula,][]{lyn62} was found to
be bright at cm-wavelengths, where no known emission mechanisms were
expected \citep[][]{fin02, fin04, cas06}. At the time of writing
LDN~1622 is the only dark cloud known to radiate at cm-wavelengths.

What is the nature of the cm-wave emitters? Do the dark clouds and the
cirrus clouds radiate by the same emission mechanisms?
\citet{dl98a,dl98b} proposed electric dipole radiation from polarized
very small dust grains (VSGs) spinning at GHz frequencies, or spinning
dust. \citet{dl99} also suggested that `magnetic dust', or magnetic
dipole emission due to thermal fluctuations in the magnetization of
ferromagnetic grains, could produce detectable cm-wave emission.
However, a finite charge density exists in atomic and molecular
clouds; a small part of the neutral material is ionised by exposure to
pervasive cosmic rays or soft-UV photons \citep[e.g.][]{tie05}. As an
alternative to spinning dust, could the residual charges radiate at
the observed levels?

Prototypical and well studied local clouds can give information on the
environments giving rise to cm-wave radiation.  The $\rho$~Oph
molecular cloud \citep[e.g.][]{enc74,you06}, at a distance $D = 135
\pm$15~pc \citep[parallax distance to HD147889,][]{hab03}, lies in the
Gould Belt of the closest molecular complexes.  $\rho$~Oph is
undergoing intermediate-mass star formation - the most massive of its
young stars is HD147889, which we show here to be a close
pre-main-sequence B2, B3 binary, not hot enough to form a conspicuous
region of ionized-hydrogen (H\,{\sc ii} region). A description of the
region can be found in Fig.~4 of \citet{you06}.

Here we present the first resolved images at cm-wavelengths of the
$\rho$~Oph main cloud, LDN~1688. We describe our observations in
Section~\ref{sec:cbiobs}, as well as auxiliary data in
Sec.~\ref{sec:aux}, including unpublished {\em ISO}-LWS parallel mode
data.  We proceed to summarise the available imaging and spectroscopic
data in Sec.~\ref{sec:obs}. The dust emission from $\rho$~Oph~W and
the contribution of spinning or magnetic dust at cm-wavelengths are
studied critically in Sec.~\ref{sec:dust}. We also propose in
Sec.~\ref{sec:CI} an alternative emission mechanism for the 31~GHz
emission based on the C\,{\sc i} continuum from a cold plasma.  In
Sec.~\ref{sec:RRLs} we analyse the non-detection of the radio
recombination line system in $\rho$~Oph~W. We discuss our findings in
Sec.~\ref{sec:disc} in terms of the C\,{\sc ii} Str\"omgren spheres
around the early type stars that are interacting with the $\rho$~Oph
cloud, before concluding in Sec.~\ref{sec:conc}.

% which we interpret as a C\,{\sc ii}
%region.

%: the
%origin of the cm-wave emission from dark clouds appears to be the
%continuum from interstellar carbon.

\section{CBI Observations} \label{sec:cbiobs}

\begin{figure}
\begin{center}
\includegraphics[width=0.8\columnwidth,height=!]{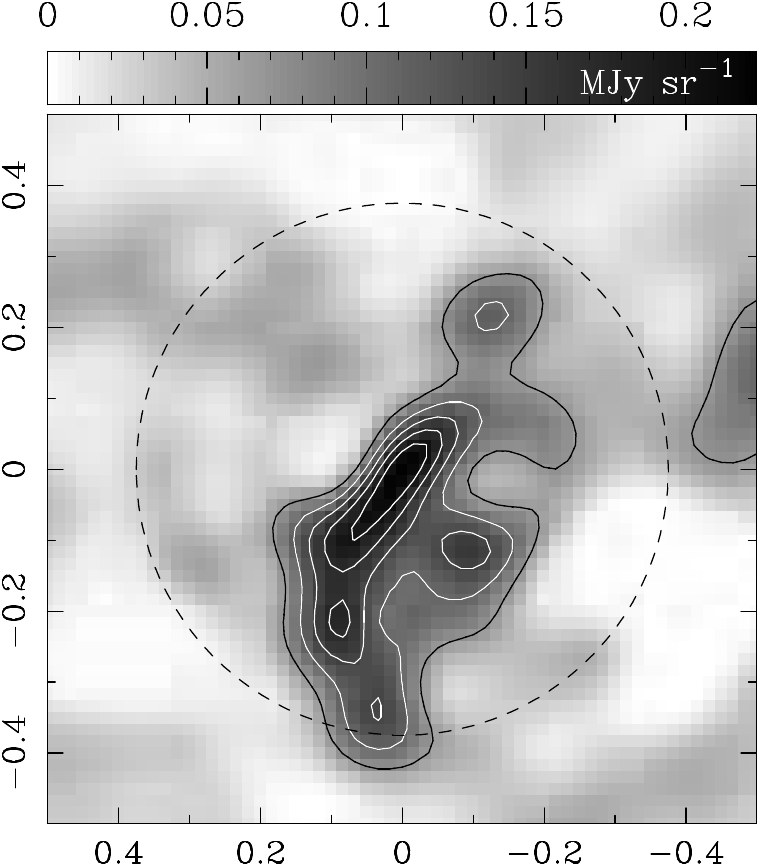}
% image_roph_CBI_1.pl
\end{center}
\caption{\label{fig:CBI0}
MEM model of the CBI 31~GHz visibilities.  $x-$ and
  $y-$axis show offset RA and DEC from $\rho$~Oph~W (J2000 16h25m57s,
-24d20m50s), in degrees of arc.  The contour levels are at 0.067
,0.107 ,0.140 ,0.170 and 0.197, in MJy~sr$^{-1}$. The dashed-circle
follows the half-maximum level of the CBI primary beam (CBI PB
hereafter).  }
\end{figure}

The 31~GHz image of $\rho$~Oph that motivates this work is shown in
Fig.~\ref{fig:CBI0}. It was reconstructed from the CBI visibilities
using a maximum entropy method. In this section we give details on the
CBI observations and image reconstruction.

\subsection{Total intensity}

The Cosmic Background Imager \citep[CBI][]{pad02} is a planar
interferometer array with 13 antennas, each 0.9~m in diameter, mounted
on a 6~m tracking platform. The CBI receivers operate in 10 frequency
channels covering 26--36~GHz.  It is located in Llano de Chajnantor,
Atacama, Chile.  With a uniform-weight synthesised beam of
$\sim$6~arcmin and a primary beam of 45.2~arcmin full-width at half
maximum at 31~GHz, the CBI is well suited to image clouds
$\sim$20--30~arcmin in total extent. The CBI primary beam (CBI PB
hereafter) encompasses most of LDN~1688.

% 7-Jul-2004, 19-Apr-2005 and 23-Apr-2005

During three nights of July 2004 and April 2005 we acquired 31~GHz
visibilities in a single pointing on the $\rho$~Oph~W
photo-dissociation region \citep[PDR,][]{hab03}, with the CBI in its
compact configuration, for a total of $\sim$12000~s on-source.
Baseline length ranged between 100~$\lambda$ and 400~$\lambda$,
corresponding to spatial scales of 34.4~arcmin and 8.6~arcmin,
respectively. Cancellation of ground and Moon contamination was
obtained by differencing with a leading reference field at the same
declination but offset in hour angle by 10~m, the duration of the
on-source integration.

\subsection{Image reconstruction}

We produced the MEM model in Fig.~\ref{fig:CBI0} by minimising the
model functional $L = \chi^2 - \lambda S$, where the expression for
$\chi^2$ is given by Eq.~A1 in \citet[][]{cas06}, and with $S= -
\sum_i I_i \log ( I_i / M_i )$, where $\{I_i\}_{i=1}^{N}$ is the model
image and $M_i$ is an image prior.  The sum extends over the number of
independent data points, which is $f=31680$ (two for each of 15840
complex visibilities). In this case we chose a regularizing parameter
$\lambda = 200$, and square model images with $256^2$ square pixels,
each 1~arcmin on a side. The optimization converged in
20~iterations. The image prior was constructed from a combination of
the IRAC~4 image at 8~$\mu$m (see Sec.~\ref{sec:spitzer}) and WMAP~Ka
at 33~GHz (see Sec.~\ref{sec:archive}). More details on image
reconstruction are given in Sec.~\ref{sec:imagereconstruction}.

\subsection{Polarization}

$\rho$~Oph was also observed  by the CBI in polarization during four
nights in August, September and October 2004.  Details of the
polarization calibration and data reduction procedures can be found in
\citet{rea04, car05, dic06, cas07}. Data from all nights were combined
together. They were mapped with DIFMAP using natural weights and
optimal noise weighting. A polarized intensity map was created using
AIPS COMB and POLC option to correct for the noise bias. The noise on
the Q and U maps is about 9 mJy~beam$^{-1}$. The beam is
9.5$\times$8.3~arcmin$^2$.

No obvious polarization signal is visible in the polarized intensity
map, and is consistent with an rms noise of 12 mJy~beam$^{-1}$. Using
the same weighting scheme (same visibilities and synthesized beam),
the peak total-intensity is 752~mJy~beam$^{-1}$.

%with a total flux density in the map of 5.6Jy (this is similar to the
%5.0Jy that Simon states in the paper from the MEM analysis).

At peak total intensity, the 1~$\sigma$ polarization limit is {\mbox
  12 / 752} =1.6\%, or a 3~$\sigma$ upper limit of 4.8\%. In the
weaker regions of $\rho$~Oph~W, where the total-intensity drops to
typically 100 -- 200~mJy~beam$^{-1}$, the polarization limit increases
to $\sim$12 / 150 = 8\%, or 24\% at 3~$\sigma$.

For the integrated flux density over a 45~arcmin diameter aperture,
the polarization limit has to be calculated using a primary-beam
corrected map which multiplies up both the signal and the noise. This
was done, making a noise corrected polarized intensity map assuming a
noise level of 45~mJy~beam$^{-1}$ (average over the field). The rms in
this image is 57~mJy~beam$^{-1}$. So the integrated polarization limit
is ~1.0\% at 3~$\sigma$.

%
%#  6: ROPHWL          16:15:57.00 -24:20:50.0  1241 1466   5248.8
%#  7: ROPHW           16:25:57.00 -24:20:50.0  1097 1168   4601.2
%
%7-Jul-2004 OK
%
%19-Apr-2005 OK
%# 25: ROPHWL          16:15:57.00 -24:20:50.0   819  949   3456.1
%# 26: ROPHW           16:25:57.00 -24:20:50.0   781  857   3275.8
%
%22-Apr-2005 OK
%
%#  5: ROPHWL          16:15:57.00 -24:20:50.0   776  943   3275.8
%#  6: ROPHW           16:25:57.00 -24:20:50.0   770  858   3229.6
%

\section{Auxiliary data} \label{sec:aux}

\subsection{{\em ISO} parallel mode data} 

\begin{figure*}
\begin{center}
\includegraphics[width=\textwidth,height=!]{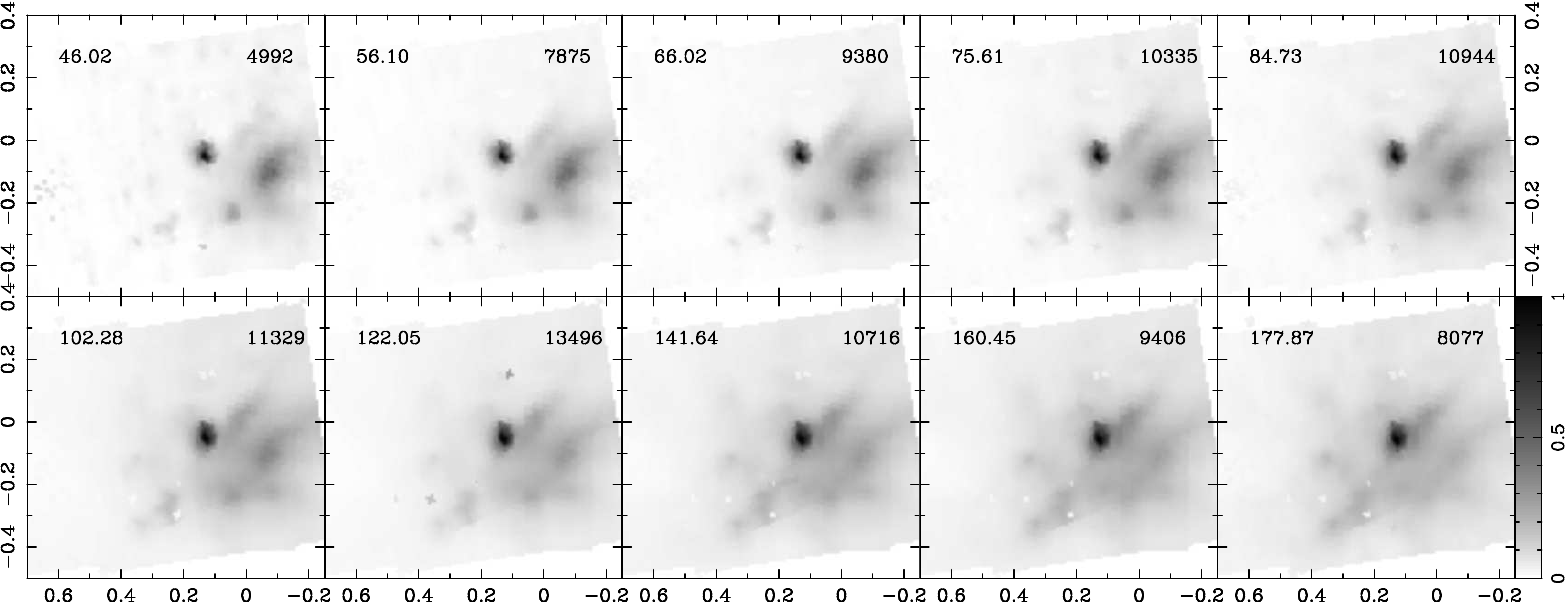}  % image_roph_global_new.pl 
\end{center}
\caption{\label{fig:iso} The {\em ISO}-LWS parallel mode data on
  $\rho$~Oph.  $x-$ and $y-$axis show offset RA and DEC from
  $\rho$~Oph~W, in degrees. The centre wavelength in microns of the
  LWS channels are indicated on the upper-left of each image. The
  intensities in MJy~sr$^{-1}$ have been scaled by the amount
  indicated on the upper-right of each image.  }
\end{figure*}

The {\em ISO} Long Wavelength Spectrograph parallel mode
survey\footnote{\tt
  http://www.iso.vilspa.esa.es\/manuals\/HANDBOOK/lws\_hb\/node1.html}
covered most of LDN~1688 in 10 bands ranging from 46~$\mu$m to
178~$\mu$m.  The full set of {\em ISO} images is shown in
Fig.~\ref{fig:iso}. The broad wavelength coverage of the {\em ISO}
data allows dust mass and extinction estimates (see
Sec.~\ref{sec:irdust}). Especially relevant to this work is the
178~$\mu$m image, which is used in the discussion of magnetic dust in
Sec.~\ref{sec:mdust}.

The parallel mode data were obtained when another ISO instrument was
operating with total sky coverage of about 1\%. The processing of
parallel data was the same as for the primary mode data. Engineering
conversions were applied to obtain the photocurrents. Once the
photocurrents were obtained the dark current was removed. The
calibration of the detector responsivity relies on a simple ratio
between the response to the illuminators found at the time of a
particular observation and that used as a reference. Parallel mode
maps were produced at ten detector wavelengths and are generated by
combining different rasters. Several data reduction tools were
developed in IDL and these form the LWS parallel interactive analysis
package that can be found at: {\tt
  http://jackal.bnsc.rl.ac.uk/isouk/lws/software/software.html}.

We have compared the {\em ISO} data at 102.26~$\mu$m with the {\em
  IRAS}~100$\mu$m survey \citep{whe91}. It is noteworthy that the {\em
  ISO}~102.26~$\mu$m intensities below 1000~MJy~sr$^{-1}$ are
typically 27~\% higher than {\em IRAS}~100~$\mu$m, 40~\% higher above
1000~MJy~sr$^{-1}$. This is in agreement with the results of
\citet{cha01} and \citet{gar01}. According to the {\em IRAS}
explanatory supplement, the gain and non-linearity of the detectors is
a function of source extension, which could be the source of the
intensity-dependent discrepancy between {\em ISO} and {\em IRAS}.

\subsection{{\em Spitzer} archival data}  \label{sec:spitzer}

The IRAC data span the range 3.6 to 8$\mu$m with four distinct bands
(centered at 3.6, 4.5, 5.6 and 8$\mu$m) and an angular resolution of
2$^{''}$. Longer wavelengths, namely 24, 70 and 160$\mu$m, are covered
by MIPS data, with angular resolutions of 9, 18 and 40$^{''}$
respectively. Both the IRAC and MIPS data considered in the present
paper come from the {\em c2d} Spitzer Legacy Survey\footnote{see the
  {\em Spitzer} Science Center press release, {\tt
    http://www.spitzer.caltech.edu/Media/releases/ssc2008-03/release.shtml}}.
The MIPS data are described by \citet{pad08}, in connection with the
young stellar population of $\rho$~Oph.

\begin{figure}
\begin{center}
\includegraphics[width=\columnwidth,height=!]{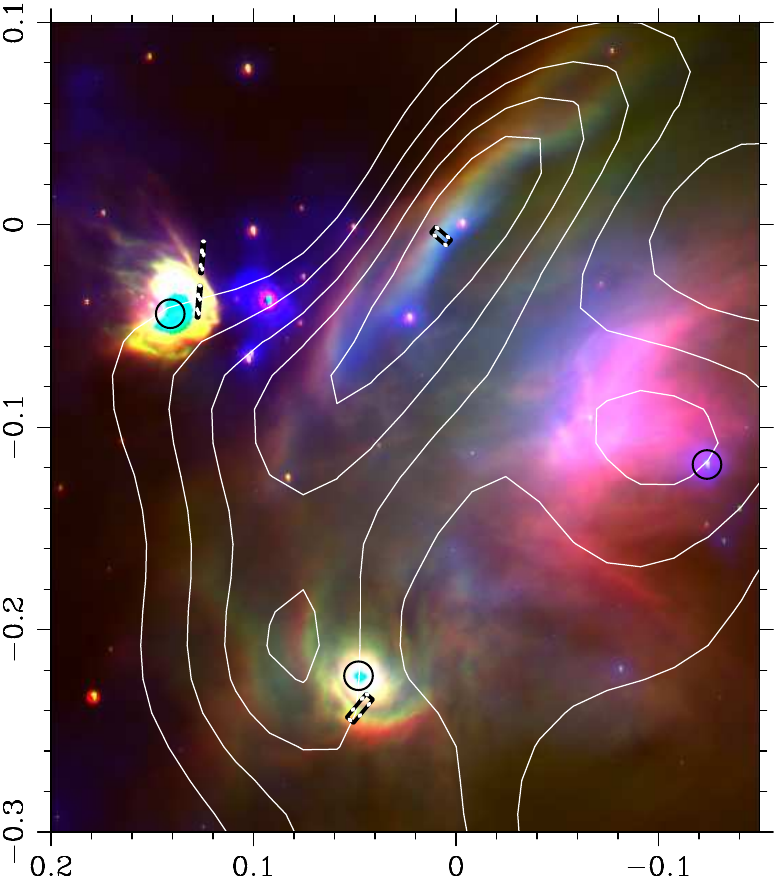}
% image_roph_S1_SR3_rgb.pl
\end{center}
\caption{\label{fig:zoom} Three-colour image of $\rho$~Oph~W: {\bf
    red}: MIPS~24$\mu$m {\bf green}: IRAC~4 at 8~$\mu$m, dominated by
  the 7.7~$\mu$m PAH band {\bf blue}: 2MASS~K$_s$-band image. $x-$ and
  $y-$axis show offset RA and DEC from $\rho$~Oph~W, in degrees. The
  dashed black and white boxes indicate the extraction apertures for
  the {\em Spitzer} IRS spectroscopy discussed in Sec.~\ref{sec:sd}
  (the apertures near S~1 are narrower than the drawing line width,
  hence they appear to be linear and not boxy). The contours follow
  the 31~GHz emission, as in Fig.~\ref{fig:CBI0}. The centre of the
  black circles indicate the positions of the early-type stars S~1,
  SR~3, and HD147889 (see Fig.~\ref{fig:multi}).}
\end{figure}

Archival {\em Spitzer} IRS data were obtained for positions near the
early-type stars S~1, SR~3 and a region of the filament. For SR3 and
the filament \citep[][]{ber07}, the data were taken in spectral
mapping mode using the high resolution (SH and LH) modules of IRS. In
this mode, the spectrograph slit is moved before each integration in
order to cover the region of interest. The Basic Calibrated Data (BCD)
files from the archive were produced using the S15 pipeline by the
{\em Spitzer} Science Centre. The spectral cubes were assembled from
the BCD files using the CUBISM software package
\citep[][]{smi04}. Spectra were extracted from the spectral cubes
using an aperture which was covered by both SH and LH modules, as
shown in Fig.~\ref{fig:zoom}. For S~1, the data were taken in
low-resolution (SL) staring mode and were also produced using the S15
pipeline. The original target was a young stellar object
\citep[J162630-242258,][]{bar97} located behind the circumstellar
nebula about S~1. We performed a full-slit extraction of SL1 with the
target in the SL1 slit and also with the target in SL2. This resulted
in SL1 spectra at two distinct positions in the S~1 nebulosity, which
we call `S~1' and `S~1off', as also shown in Fig.~\ref{fig:zoom}. The
position `S~1off' is located at the northern edge of the IR nebulosity
surrounding the early-type star S~1.

\subsection{FEROS and UNSW-MOPS observations}

\subsubsection{FEROS} \label{sec:feros}

HD147889 is the main source of excitation in $\rho$~Oph~W so an
accurate measurements of its spectral type is necessary.  HD147889 is
catalogued in SIMBAD as a single B2III/IV star \citep{hou88}.  However
\citet{haf95} reported that HD147889 is a close binary.  The spectral
types they quote are B2 for both components, and luminosity class
IV-V. Unfortunately \citet{haf95} do not give details on their
atmospheric models.  Given the importance of HD147889 for $\rho$~Oph~W
we undertook new echelle observations and up-to-date atmospheric
modelling.

In order to precisely determine the UV field impinging on $\rho$~Oph~W
we acquired echelle spectroscopy of HD147889 during 2 contiguous
nights in May 2008 (ESO programme 081.C-2003). We used the FEROS
echelle spectrograph, at the ESO~2.2~m telescope, which provides full
coverage over 3500--9000~\AA~ at $R = 50000$. We also found two
spectra from Feb. 2006 in the ESO archive (programme 076.C-0164).

\subsubsection{UNSW-MOPS} \label{sec:mops}

The UNSW-MOPS\footnote{{\tt
    http://www.narrabri.atnf.csiro.au/mopra/mops/}} spectrometer on
the Mopra telescope, a 22~m single-dish, allows the acquisition of
area-spectroscopy data cubes up to $\sim$0.6~km~s$^{-1}$ in spectral
resolution and with a bandwidth of $\sim 2~10^3$~km~s$^{-1}$ (in zoom
mode).

Radio recombination lines (RRLs) are diagnostics of physical
conditions.  In particular low-frequency carbon RRLs have been
reported from $\rho$~Oph and other PDRs in reflection nebulae
\citep{pan78}.  We attempted to detect the high-frequency carbon RRL
system of $\rho$~Oph~W using UNSW-MOPS.  We acquired $20\times
20$~arcmin on-the-fly scans centred on $\rho$~Oph~W and the following
rest-frame frequencies (in GHz): in the K band, C65$\alpha$ 23.41595,
C66$\alpha$ 22.37532, C67$\alpha$ 21.39545, C68$\alpha$ 20.47197,
C69$\alpha$ 19.60089, C70$\alpha$ 18.77853, C71$\alpha$ 18.00153,
C72$\alpha$ 17.26682, C73$\alpha$ 16.57156, and in the W band,
C42$\alpha$ 85.73114 and C43$\alpha$ 79.95252. All 9 K-band
frequencies could be mapped simultaneously in zoom mode, for a total
of 40~mn.  The W-band map represented 2~h.

Data reduction was carried out with the ``livedata'' and ``gridzilla''
packages. We chose to resample the data cubes into 5~arcmin square
pixels. The 1~$\sigma$ noise was 0.366~MJy~sr$^{-1}$ (or 43~mK) for
C73$\alpha$, and 188~MJy~sr$^{-1}$ (or 832mK) for C42$\alpha$. No
lines were detected.

\subsection{Additional archive data}   \label{sec:wmap}

The {\em WMAP} satellite \citep[][]{hin07} provides low-resolution
images of $\rho$~Oph in 5 bands at 23, 33, 41, 61, and 94~GHz, with
average beam-widths of 0.88, 0.66, 0.51, 0.35, and 0.22~deg,
respectively.  Despite their low resolution, the {\em WMAP} data allow
the extraction of the radio spectral energy distribution (SED).

\subsection{Additional archive data}   \label{sec:archive}

The Parkes-MIT-NRAO survey at 5~GHz traces H\,{\sc i} free-free
emission from H\,{\sc ii} regions \citep[hereafter PMN survey,][as
  presented in {\em SkyView}, {\tt
    http://skyview.gsfc.nasa.gov}]{co93}. The PMN survey at 5~GHz is
strongly affected by flux loss, or missing low spatial frequency due
to high-pass filtering. The $\rho$~Oph~W region includes extended
negatives, indicative of survey artifacts. We estimated the level of
filtering artifacts in PMN by comparing with diffuse H\,{\sc ii}
regions \citep[selected from][ hereafter LPH96]{loc96} in the
Effelsberg 2.7~GHz survey \citep[][]{rei90}.  We extract flux
densities as in $\rho$~Oph~W (i.e. using a circular aperture 45~arcmin
indiameter), and scale with a free-free index.  The flux density
recovered by PMN is 87\% in the case of relatively compact RCW~6
($\sim$6~arcmin), 25\% in LPH96~201.663+1.643 \citep[$\sim$15~arcmin,
  see also the analysis in][]{dic06}, and only 6\% in
LPH96~78.229+3.716 (a filament $\sim$1~deg long and $\sim$15~arcmin
wide). It appears that the flux recovered by PMN in the case of
$\rho$~Oph~W could be as low as 5--10~\%.

Another probe of the diffuse emission in $\rho$~Oph is the 2MASS
survey {\tt http://www.ipac.caltech.edu/2mass/}. The 2MASS-K$_s$ image
produced by the Montage\footnote{\tt http://montage.ipac.caltech.edu/}
mosaicing software, shown in Fig.~\ref{fig:zoom} and \ref{fig:2MASS},
is an interesting comparison point with the CBI image (see
Sec.~\ref{sec:morph} and Sec.~\ref{sec:rovibH2}).

Our analysis also makes use of the Southern H-Alpha Sky Survey Atlas
\citep[SHASSA][]{gau01}. Another map useful in constraining the radio
properties of $\rho$~Oph is that of \citet{baa80}, who report a
2.3~GHz map of the $\rho$~Oph region, with a beam of 20~arcmin FWHM.

\section{Observed properties} \label{sec:obs}

\subsection{Morphology} \label{sec:morph}

\begin{figure*}
\begin{center}
\includegraphics[width=\textwidth,height=!]{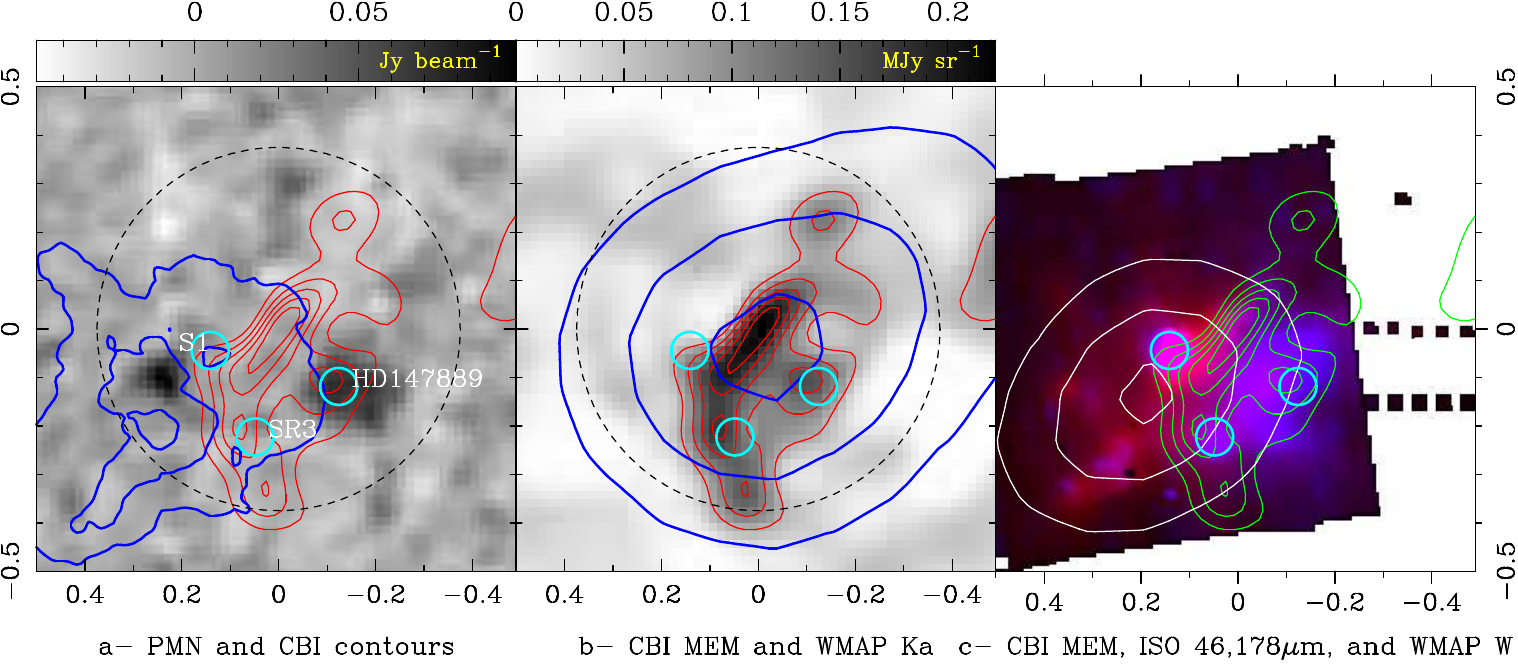}  % image_roph_global_new.pl 
\end{center}
\caption{\label{fig:multi} Morphological evidence ruling out standard
  dust or optically thin H\,{\sc i} free-free as the origin of the
  31~GHz emission.  $x-$ and $y-$axis show offset RA and DEC from
  $\rho$~Oph~W, in degrees.  The centre of the cyan circles indicate
  the positions of the early-type stars S~1, SR~3, and HD147889. {\bf
    a-} CBI MEM contours overlaid on the PMN image (4.85~GHz), with
  $A_\mathrm{V}$ in blue contours at 10 and 20. {\bf b-} The CBI MEM
  image is shown in grey scale in MJy~sr$^{-1}$, with the same contour
  levels as in Fig.~\ref{fig:CBI0}.  {\em WMAP}~33~GHz is shown in blue
  contours at 2.52, 2.99, and 3.37~K.  {\bf c-} CBI MEM contours
  overlaid on an {\em ISO} colour map with 46$\mu$m in blue and
  178$\mu$m in red.  {\em WMAP}~94~GHz is shown in white contours at
  1.21, 1.51, and 1.75~K.  }
\end{figure*}

The most conspicuous feature in the CBI MEM model, also shown in
Fig.~\ref{fig:multi}~b, is the $\rho$~Oph~W PDR, at the origin of
coordinates.  The {\em WMAP}~33~GHz contours in Fig.~\ref{fig:multi}b
confirm the bulk morphology of the CBI image.

We indicate in Fig.~\ref{fig:multi}a the positions of the early-type
stars S~1 \citep{gra73}, SR~3 \citep[][{\tt EM* SR 3} in SIMBAD, also
  known as Elia 2-16]{eli78}, and HD147889. These three stars,
undetected in the CBI maps, excite the region and serve as reference
points.

Inspection of Fig.~\ref{fig:multi}c leads to the conclusion that the
31~GHz emission from $\rho$~Oph is not the Rayleigh-Jeans tail of the
sub-mm-emitting large dust grains (``standard dust'' hereafter).  If
the 31~GHz emission was due to standard dust, it should follow 94~GHz
and 178~$\mu$m. Both ISO~178~$\mu$m (in red in Fig.~\ref{fig:multi}c)
and {\em WMAP}~94~GHz trace standard dust. But the CBI contours are
offset to the North-West. The peak in {\em WMAP}~94~GHz is separated
by 14~arcmin from the peak in the CBI model. The 94~GHz dust has been
resolved into a system of cold dust clumps by sub-mm bolometer arrays
\citep[e.g.,][]{you06,rid06}.

%the {\sc COMPLETE} database on$\rho$~Oph][]{rid06}.
%http://www.cfa.harvard.edu/COMPLETE/

%In Fig.~\ref{fig:multi}c we report the restored CBI image, produced by
%adding the primary-beam-corrected dirty-map residuals to the
%convolution of the model image and the natural-weight synthetic beam.

The PMN image in Fig.~\ref{fig:multi}a shows two features: extended
emission about HD147889, which having an H$\alpha$ counterpart is a
faint H\,{\sc ii} region, and the point source LFAM 21
\citep[J162700.0-242640*, $\sim$60~mJy at 5GHz,][]{gag04}, undetected
at higher frequencies. Also shown in Fig.~\ref{fig:multi}a are
contours at 10 and 20 of the visual extinction map $A_\mathrm{V}$ from
\citet[][]{rid06}\footnote{{\tt
    http://www.cfa.harvard.edu/COMPLETE}}. $A_\mathrm{V}$ traces the
total column density of material, i.e.  the total mass density of the
cloud.

%\begin{figure}
%\begin{center}
%\includegraphics[width=\columnwidth,height=!]{CBI_VIR_roph.pdf}
%\end{center}
%\caption{\label{fig:CBI} CBI image of $\rho$~Oph~A. $x-$ and $y-$axis
%  show offset RA and DEC from $\rho$~Oph~W (J2000 16h25m57s,
%  -24d20m50s). The grey scale shows the MEM model of the CBI data, in
%  MJy~sr$^{-1}$, with countour levels at 0.040, 0.080, 0.120, and
%  0.159. The thick solid line are contours of the restored CBI image,
%  with contours at 0.165 and 0.487 Jy~beam$^{-1}$, and the CBI
%  synthetic beam is shown on the upper right.}
%\end{figure}

%\caption{
%
%%\label{fig:multi} 
%
%%The CBI image of $\rho$~Oph~A and
%%  comparison templates.  $x-$ and $y-$axis show offset RA and DEC from
%%  $\rho$~Oph~W (J2000 16h25m57s, -24d20m50s). 
%%
%%${\bf a -}$ 
%
%% The grey scale in the CBI image shows a MEM model, in MJy~sr$^{-1}$,
%% with countour levels at 0.040, 0.080, 0.120, and 0.159. The CBI
%% synthetic beam is shown on the upper right of the CBI restored
%% image.
%%
%
%
%% We use contours from the CBI restored image in the comparison
%% with WMAP Ka and W (grey-scale units in Jy), and the CBI MEM model
%% for comparison with IRIS~12$\mu$m and PMN. We indicate the the centre
%% coordinates of the $\rho$~Oph~A, B and W clouds, and the stars S~1,
%% SR~3 and HD147889.  
%%
%}
%

Diffuse mid-IR emission ($<$60~$\mu$m) from $\rho$~Oph is interpreted
as stochastic heating of VSGs \citep[][]{ber93}. In the dust model of
\citet{dl07} VSGs are regarded as large polycyclic aromatic
hydrocarbons (PAHs). Only the smallest VSGs, or PAHs, can reach
spinning frequencies of $\sim$30~GHz. Thus for spinning dust a close
correspondence is expected between the CBI and mid-IR templates. There
are indeed similarities between 31~GHz and {\em ISO}~46$\mu$m (in blue
in Fig.~\ref{fig:multi}c): both are found near HD147889, while the
far-IR emission is located about S~1 and to the south-east.

But there are important differences between the CBI image and the IR
templates.  In Fig.~\ref{fig:zoom} we compare near- and mid-IR maps
with the CBI contours. IRAC and MIPS aboard {\em Spitzer} provided
high resolution mosaics of $\rho$~Oph at near- and far-IR
wavelengths. The dust-heating flux from S~1 produces the most
conspicuous mid- and far-IR nebula in the entire $\rho$~Oph region
\citep[e.g. the PAH~6.7~$\mu$m band image in ][]{abe96}. Yet neither
S~1 nor SR~3 have 31~GHz counterparts.  It can also be noted that the
photospheric IR emission is absent in the CBI maps \citep[as is the
  case in LDN~1622,][]{cas06}.

% even though
%Fig.~\ref{fig:CBI}e clearly shows that the MEM models pick up the
%infrared nebulae about S~1 and SR~3.

%The strongest diffuse mid-IR emission stems from the circumstellar
%nebulae about the embedded early-type stars S~1 and SR~3, neither of
%which have 31~GHz counterparts.

%After simulation of the CBI
%beam, the peak intensity in the IRAC~8$\mu$m nebulosity about HD147889
%is 1.2 times the peak intensity in $\rho$~Oph~W, while this ratio is
%0.7 in the CBI data \footnote{We have cross-checked by simulating CBI
%  observations on the IRAC~8$\mu$m image that the mid-IR nebula about
%  HD147889 is sufficiently compact not to be filtered-out in the
%  interferometer data}
%

%Interferometric data is insensitive to the lowest spatial frequencies.

%%%%%%%%%%%%%%%%%%%%%%%%%5

\begin{figure}
\begin{center}
\includegraphics[width=\columnwidth,height=!]{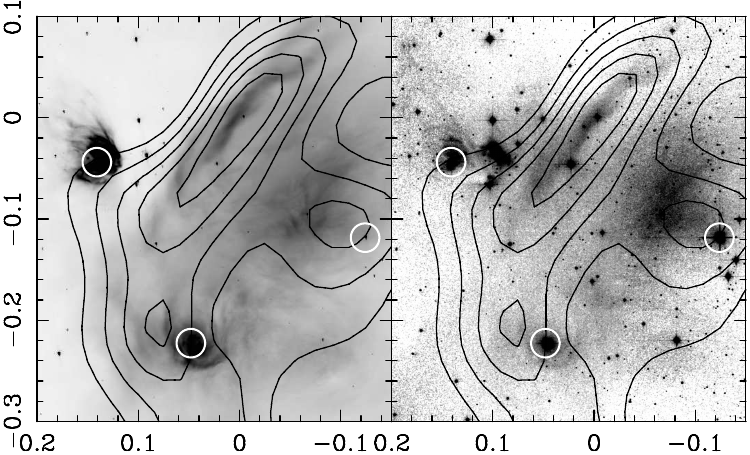}
% image_roph_S1_SR3_2MASSgrey.pl
\end{center}
\caption{\label{fig:2MASS} The 2MASS-CBI correlation. The image is an
  overlay of the CBI MEM contours on the IRAC~8~$\mu$m (left) and
  2MASS~K$_s$-band images (right). $x-$ and $y-$axis show offset RA and
  DEC from $\rho$~Oph~W, in degrees. It may be appreciated that the
  2MASS and IRAC~8~$\mu$m emissions correlate with 31~GHz in
  $\rho$~Oph~W.}
\end{figure}

The IR emission shifts to longer wavelengths radially away from
HD147789: at increasing wavelengths the emission from $\rho$~Oph~W
moves to the N-E. It is surprising that the best match to the CBI
contours of $\rho$~Oph~W turns out to be the 2.2~$\mu$m diffuse
emission in the 2MASS~K$_s$-band, shown in blue in Fig.~\ref{fig:zoom},
and in grey scale in Fig.~\ref{fig:2MASS}.  The pointing uncertainty
of the CBI is $<0.5$~arcmin \citep[e.g.][]{cas06} - the phase
calibrator J1626-298 was offset by only 0.2~arcmin from the phase
centre. Thus the CBI data are sensitive to the $\sim$2~arcmin
translation in $\rho$~Oph~W, when seen in MIPS~24$\mu$m and
2MASS~K$_s$-band.

The only 2MASS counterpart to the CBI contours are the emission peaks
along $\rho$~Oph~W. The north-south extension of the CBI contours to
the east of SR~3 is undetectable in 2MASS, but we note that this
region has higher extinction than $\rho$~Oph~W (see
Sec.~\ref{sec:irdust}).

Another conspicuous feature of the 2MASS image is the nebulosity about
HD147889, which is probably the near-IR counterpart to the diffuse
H\,{\sc ii} region seen in PMN. An extended 2MASS source at
J162622-242301, at ($\Delta \alpha, \Delta \beta ) \approx ( +0.09,
-0.04 )$ and between S~1 and W, corresponds to the disk-like envelope
surrounding the young stellar object GSS~30~IRS~1 \citep[][]{chr96}.

\subsection{Spectrum} \label{sec:spec}

\begin{figure}
\begin{center}  % SED_ROPH.pl
\includegraphics[width=\columnwidth,height=!]{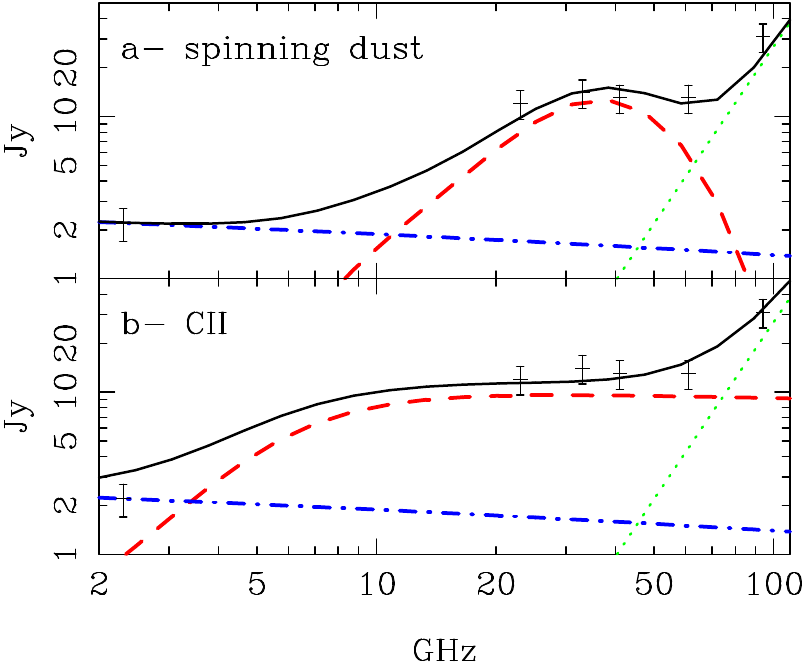}
\end{center}
\caption{\label{fig:spec} The cm-wave spectrum in a 45.5~arcmin
  circular aperture centred on $\rho$~Oph~W. All data points are
  extracted from the {\em WMAP} images, except for the 2.3~GHz point,
  which is taken from \citet[][see Sec.~\ref{sec:HIff}]{baa80}. The
  solid black line is a fit to the data that includes a modified
  blackbody, in green dotted line (with $T_d = 23~$K and $\beta =
  1.7$), a diffuse free-free component, shown as a dash-dotted blue
  line, and in red-dashed line either a spinning dust component ({\bf
    a- spinning dust}, for the `DC' environment), or a cold C\,{\sc i}
  thermal component stemming from an ensemble of PDR sheets seen
  edge-on ({\bf b- C\,{\sc ii}}).}
\end{figure}

In Fig.~\ref{fig:spec} we have constructed the spectral energy
distribution of $\rho$~Oph~A using the flux densities given in
Table~\ref{table:SED}. The {\em WMAP} flux densities are extracted
from a circular aperture equal to the CBI primary beam, without
background correction. 

%The 2.3~GHz point, taken from \citet{baa80}, is
%discussed in Sec.~\ref{sec:HIff}. 

%We avoided the use of the CBI flux
%densities because they are affected by flux-loss.  

%We plot flux density in Jy
%  ($y-$axis) as a function of frequency in GHz ($x-$axis).  

\begin{table}
\centering
\caption{Observed flux densities inside the half-maximum contour of
  the CBI primary beam. Root-mean-square uncertaintes are of order
  20\%.}
\label{table:SED}
\begin{tabular}{llllllll}
\hline 
$\nu^a$     &   2.3$^c$   &  23$^d$  & 31$^e$  & 33$^d$ & 41$^d$ & 61$^d$  & 94$^d$  \\ 
$F(\nu)^b$  &   2.2       &  12      & 7.9      &  14    & 13     &  13 & 31  \\ \hline 
\end{tabular}
\begin{flushleft}
$^a$  Frequency in GHz.
$^b$  Flux density in Jy, with uncertainties of 20\%. 
$^c$  From \citet{baa80}. 
$^d$  From {\em WMAP}
$^e$  CBI flux density before correcting for flux loss.
\end{flushleft}
\end{table}

It is difficult to extract spectral index information from the 10 CBI
channels because the $u,v$ coverage varies with frequency: flux loss,
or the fraction of the flux density in the CBI 45~arcmin aperture that
is filterout out, varies with both channel frequency and sky image. A
template close in morphology to the 31~GHz emission could have been
used to extract accurate channel by channel flux densities. But no
such template is available.  The cross-correlation between the CBI
data and simulated visibilities on the prior image shown in
Fig.~\ref{fig:CBI}e gave high reduced $\chi^2$ values ($\sim 2.4$) and
$r$ correlation coefficients of $\sim 0.6$.  The MEM model itself
cannot be used as a template in this case because it is limited in
resolution by the dynamic range of the data.

The sub-mm continuum in $\rho$~Oph~A can be estimated from the {\em
  WMAP}~94~GHz and far-IR points, as a single modified black body with
an emissivity index $\beta$.  For this purpose we used the IRIS
\citep[][]{miv05} reprocessing of the {\em IRAS} survey \citep{whe91},
and extracted flux densities in a circular aperture equal to the CBI
primary beam (the {\em ISO} images lack coverage in the Eastern side
of the photometric aperture).  The observation of a sub-mm flux
density is required to constrain the value of $\beta$. We did not use
the SCUBA maps of $\rho$~Oph because they are strongly filtered. We
estimated by comparison with the expected sub-mm continuum that the
diffuse emission that is filtered out of the bolometer data in a
45~arcmin aperture can amount to $\sim$85\%.

%Any value of $\beta$, within 1.5 and
%2.3, can fit the far-IR points. 

\subsection{CBI flux loss}  \label{sec:cbiloss}

% WMAP beams :  0.88 	0.66 	0.51

The CBI image is not sensitive to the lowest spatial frequencies;
uncertainties in the comparison data are approximate and stem mostly
from mismatched angular scales, not thermal noise. In particular the
{\em WMAP} beam at 33~GHz is approximately Gaussian with
FWHM$\sim$49~arcmin \citep{ben03}, so that the emission within the CBI
beam is somewhat convolved with the surroundings. In turn, part of the
flux at low spatial frequencies is filtered out in the interferometer
data.

We have simulated CBI observations on two templates: IRAC~8$\mu$m,
after removing stars by median-filtering, and IRIS~1, a 12$\mu$m
template \citep[][]{miv05}. After renconstruction with a blank prior
(as in Sec.~\ref{sec:imagereconstruction}), we find that 41.3\% of the
IRAC~8$\mu$m flux within the CBI primary beam is recovered by the
simulation; and 51.4\% in the case of IRIS~1. The same algorithm
applied to the CBI visibility produces the image shown on
Fig.~\ref{fig:CBI}d, and a flux density of 5.0~Jy, which when compared
to 14~Jy in {\em WMAP}~33~GHz implies that the fraction of flux
density recovered in the blank-prior reconstructions is 35\%. But the
image prior used in our best CBI reconstructions allow recovering part
of the extended emission. Comparison of the flux density obtained from
the restored image on Fig.~\ref{fig:CBI}a and with {\em WMAP}~33~GHz
shows that, with the use of a prior, 56\% of the signal is recovered
by the CBI.

%In the absence of better knowledge, we adopted
%conservatively large 1~$\sigma$ uncertainties of 20\% for all data
%points.

\subsection{HI free-free}  \label{sec:HIff}

\begin{figure}
\begin{center}
\includegraphics[width=0.8\columnwidth,height=!]{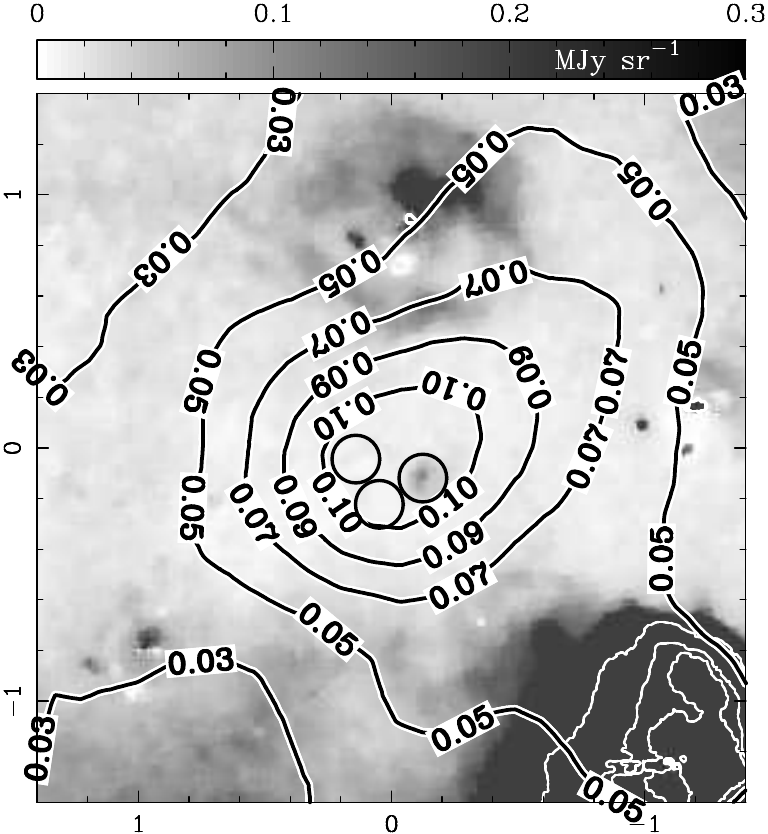}
% image_roph_WMAP_SHASSA.pl
\end{center}
\caption{\label{fig:shassa_wmap} H$\alpha$ and H\,{\sc i} continuum
  towards $\rho$~Oph. The Fig. shows an H$\alpha$+[N\,{\sc ii}] image
  of the $\rho$~Oph region, as extracted from SHASSA, and scaled to
  the corresponding $T_e = 8000~$K free-free continuum at 33~GHz, in
  MJy~sr$^{-1}$, without correction for reddening. The black and white
  contours follow {\em WMAP}~33~GHz, also in MJy~sr$^{-1}$. The SHASSA
  emission from Gum~65 to the south-west is traced in white contours
  at 1.0, 5.0, and 10~MJy~sr$^{-1}$.  The circles indicate the
  positions of S~1, SR~3, and HD147789. $x-$ and $y-$axis show offset
  RA and DEC from $\rho$~Oph~W, in degrees. }
\end{figure}

The {\em WMAP}~33~GHz and SHASSA images shown in
Fig.~\ref{fig:shassa_wmap} illustrate that $\rho$~Oph~W is superposed
on an extended background, which is probably the free-free counterpart
to the H$\alpha$ emission. There are however intriguing features in
Fig.~\ref{fig:shassa_wmap}: the bright H$\alpha$ source to the
south-west (a bright H\,{\sc ii} region surrounding $\sigma$~Sco,
known as Gum~65 or S~9) should correspond to 33~GHz free-free levels
at least a factor of ten higher than seen in {\em WMAP}~Ka. We believe
this is due in part to dillution in the {\em WMAP} beam, but also to
[N\,{\sc ii}] doublet contamination in the SHASSA filter, which can be
of order 50\% in photoionised nebulae \citep[e.g. in the Helix, see
  Sec.~2.3 in][]{cas04}, or even higher in shocks. Nitrogen is found
singly ionised in the outer bounds of ionisation-bounded nebulae,
which are the surface regions sampled by the (extinction sensitive)
SHASSA filter.

In the 2.3~GHz map of \citet{baa80} the CBI primary beam lies in a
region of uniform emission, with a brightness temperature of
$\sim$100~mK. The integrated flux density inside our photometric
aperture is thus $\sim$2.2~Jy at 2.3~GHz, or $\sim$1.7~Jy at
31~GHz\footnote{Given that the CBI flux density measured from the
  blank-prior CBI image in Fig.~{fig:CBI}d is 5.0~Jy while {\em
    WMAP}~33GHz sees 14~Jy, of which 1.7~Jy are diffuse free-free, we
  obtain that the CBI recovers 40\% of the non-free-free emission
  (consistent with Sec.~\ref{sec:cbiloss}, see also
  Sec.~\ref{sec:discconc})}.

%***** Cross check , WMAP gives a
%background of 4 Jy inside the 45armcin aperutre, if at
%0.03MJy/sr*****. 

The PMN flux density in the 45~arcmin photometric aperture is
essentially zero. Yet the SHASSA survey shows substantial diffuse
H$\alpha$ in the region of $\rho$~Oph~W.  The total H$\alpha$ flux in
the CBI primary beam, if due to the diffuse 8000~K plasma discussed by
\citet[][]{dic03}, implies that the free-free level at 5~GHz should be
at least $\sim$2.5~Jy. After correction for the average value of
extinction inside the CBI beam but outside the dark cloud itself,
$E(\mathrm{B}-\mathrm{V})\approx 1-3$ \citep[][]{sch98}, or
$A_\mathrm{V} = 3-9$ for $R_\mathrm{V} = A_\mathrm{V} /
E(\mathrm{B}-\mathrm{V}) = 3.1$, the predicted level of free-free
emission could reach values of order $\sim$10Jy. Since the 2.3~GHz
flux density inferred from \citet{baa80} is somewhat less than the
minimum allowed by SHASSA, we conclude that the diffuse emission seen
by SHASSA is also strongly contaminated by [N\,{\sc ii}].

%We can estimate the magnitude of the diffuse free-free background in
%the CBI PB with the median intensity in a 50~arcmin wide ring
%surrounding our extraction aperture. We find that the flux density of
%diffuse free-free inside the CBI PB should be $\sim$7.2~Jy at 31~GHz
%(or $\sim$8.6~Jy at 5~GHz), so that the CBI sees about 75\% of the
%remainder 6.7~Jy at 31~GHz. This is an indication that the 31~GHz
%emission from $\rho$~Oph~W itself should be rather compact - more so
%than the IR emission, where the flux recovered by the CBI is only
%35--40\%.
%

% ~/ROPH/SED_ROPH.pl

%, where we also indicate by a thin red line the average contribution
%  of C\,{\sc i} radio recombination lines to broad-band data.

\section{Dust emission} \label{sec:dust}

\subsection{IR emission from dust grains}  \label{sec:irdust}

Bulk physical properties of the dust in $\rho$~Oph can be obtained by
fitting modified black bodies to the spectra extracted from each
spatial pixel of the {\em ISO} data cube. A single modified black body
turns out to give a fairly good fit, despite expected temperature
gradients, especially from stochastic heating of the smaller grains
(whose emission, as traced by IRAC~8~$\mu$m for instance, is very
different from that of the larger grains emitting at 100~$\mu$m).

\begin{figure}
\begin{center}
\includegraphics[width=\columnwidth,height=!]{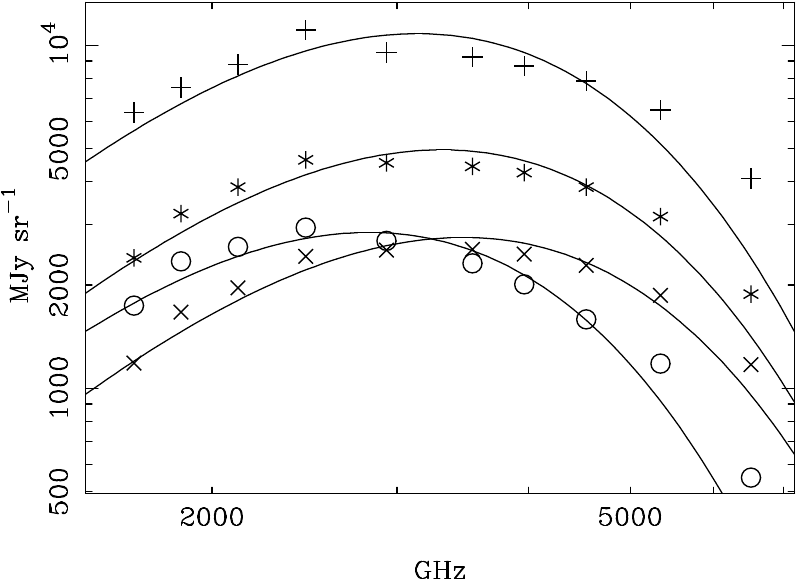}
% image_roph_irdust.pl
\end{center}
\caption{\label{fig:ISOSEDs} Representative ISO LWS specific intensity
  spectra, at four selected positions in $\rho$~Oph: $\rho$~Oph~W
  (circles), S~1 (pluses), HD~147889 (crosses) and SR~3 (asteriks,
  scaled by a factor 1.8 for clarity). The solid lines are best-fit
  grey-body spectra, as discussed in the text and summarised on
  Fig.~\ref{fig:NT}.}
\end{figure}

The grey-body parameters are the opacity at 100~$\mu$m, $\tau_{100}$,
and $T_d$, the dust temperature.  The dust emissivity index is
determined by its anticorrelation with dust temperature, as reported
by \citet{dup03}: $\beta=1/(0.4+0.008 T_d)$.  $\tau_{100}$ is related
to $N_H$, the H-nucleus column density, by $N_H = 4~10^{24} (\lambda /
100~\mu\mathrm{m})^\beta ~ \tau_{100}$~cm$^{-2}$
\citep{dra99b}. Example infrared spectra are shown on
Fig.~\ref{fig:ISOSEDs}.

\begin{figure}
\begin{center}
\includegraphics[width=\columnwidth,height=!]{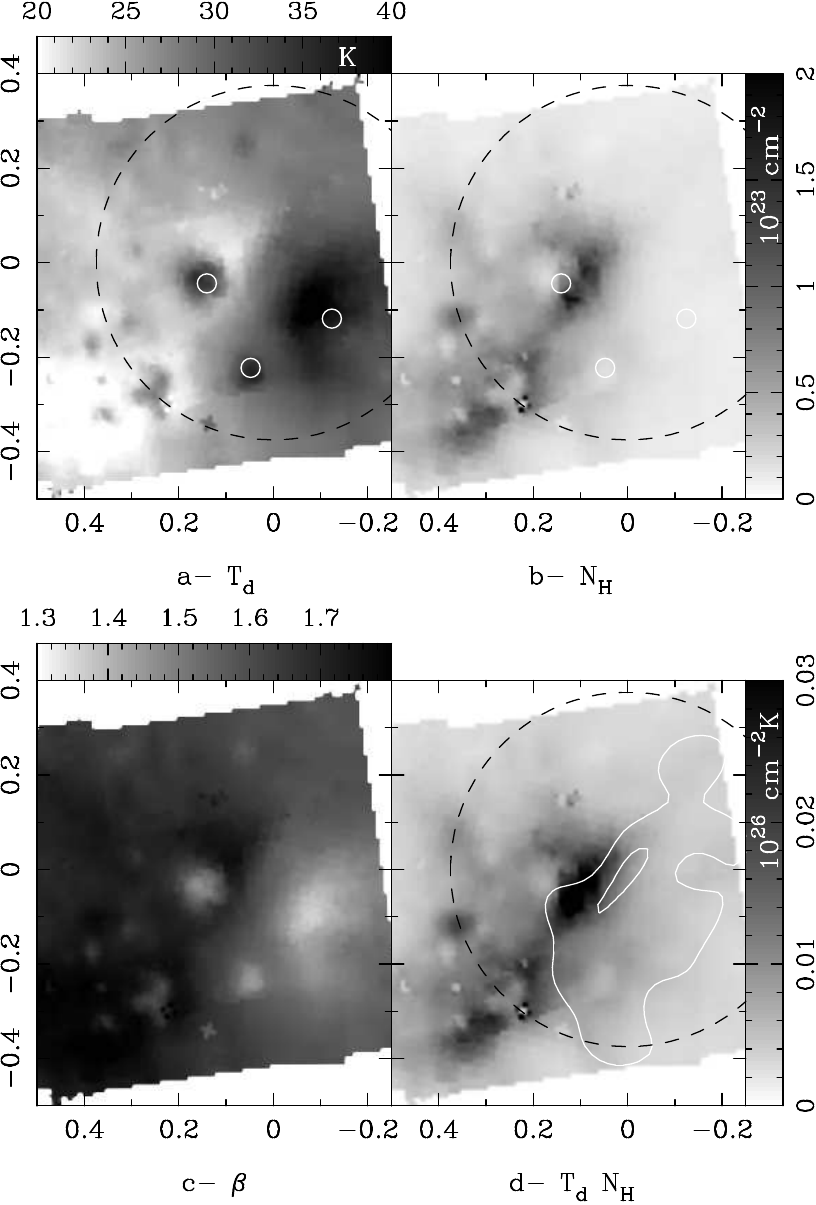}
% image_roph_irdust.pl
\end{center}
\caption{\label{fig:NT} Axes labels and symbols follow from
  Fig.~\ref{fig:multi}. {\bf a-} Dust temperature in gray scale, the
  CBI primary beam and the positions of S~1, SR~3 and HD147789 are
  indicated. {\bf b-} The H-nucleus column density inferred from the
  far-IR {\em ISO} data. {\bf c-} Dust emissivity (or temperature
  spectral index) inferred from the $\beta$-$T_d$ anticorrelation of
  \citet{dup03}. {\bf d-} Gray scale image of the product of the
  H-nucleus column densty $N_H$ times the dust temperature $T_d$, in
  units of $10^{26}$~m$^{-2}$~K. Also shown are the lowest and highest
  contours of the CBI MEM models, at 0.067 and 0.197~MJy~sr$^{-1}$.  }
\end{figure}

The resulting temperatures, column densities and dust emissivities are
shown in Figures~\ref{fig:NT}a,~\ref{fig:NT}b and ~\ref{fig:NT}c. The
maximum value of $N_H$ in Fig.~\ref{fig:NT}b is $2.3~10^{23}$
cm$^{-2}$, and the temperature varies over 20--40~K. The early type
stars S~1, SR~3 (see Sec.~\ref{sec:stars}) are coincident with
temperature peaks, as expected from grains in radiative equilibrium
with the stellar UV radiation. But neither S~1 nor SR~3 correspond to
peaks in column density. $N_H$ peaks close to S~1, but its morphology
follows that of the $\rho$~Oph~A molecular core. Therefore the IR
circumstellar nebulae about S~1 and SR~3 are radiation-bounded rather
than matter-bounded.

The total mass in the {\em ISO} field, $M_T = m_p D^2 \int d\Omega
N_H$, is 922~M$_\odot$. Considering the numerous assumptions implicit
in converting dust emissivities to H-nucleus densities, this mass
estimate is consistently close to $\sim$~2500~M$_\odot$ given by
\citet{lis95,lis99}.  But the mass seen by {\em ISO} is much less than
the 7400~M$_\odot$ derived from the near-IR extinction
\citep{rid06}. Perhaps the extinction has been overestimated, or there
are very cold regions in $\rho$~Oph not sampled by {\em ISO}.

We also obtain a mass of 65~M$_\odot$ for $\rho$~Oph~W, if its
extension is defined by a 31~GHz intensity threshold of
0.124~MJy~sr$^{-1}$ in the MEM model (see Fig.~\ref{fig:CBI0}).  The
mass above a 31~GHz intensity threshold of $10^{-3}$~MJy~sr$^{-1}$) is
359~M$_\odot$.

\subsection{Magnetic dust} \label{sec:mdust}

The magnetic dipole emission from magnetic fluctuations in
ferromagnetic grains results in a cm-wave enhancement of the grain
emissivity \citep{dl99}. At cm-wavelengths classical (not VSG) dust is
in the Rayleigh-Jeans regime, so that magnetic dust emission should be
proportional to the product of the column of ferromagnetic grains and
their temperature. Under the assumption that ferromagnetic grains, if
they exists, are uniformly mixed with the rest of the dust, the CBI
image should be proportional to the product of the dust column $N_d$
and temperature $T_d$ for constant dust-to-gas ratio ($N_d \propto
N_H$).

It can be seen in Fig.~\ref{fig:NT}d that the morphology of the CBI
MEM contours is very different from that of the product $N_H T_d$,
which instead follows the far-IR emission. Since we cannot find any
reason why ferromagnetic grains, if they exist, would be found
preferentially in $\rho$~Oph~W, we conclude that magnetic dust is
inconsistent with the {\em ISO} data.

Another difficulty for magnetic dust is the lack of detectable
polarization. \citet[][their Fig.~9]{dl99} predict polarization
fractions $f$ from perfectly aligned ferrogmanetic grains.  For their
hypothetical material 'X4' the predicted fraction at 31~GHz is at
least $f = 25\%$. X4 is the only material considered by \citet{dl99}
whose emissivities reach the levels required by the anomalous
foreground. \citet[][his Table~2]{mar07} quantifies the degree of
grain alignment through a reduction factor $R$, which ranges from 0.19
to 0.67 for a wide variety of grain geometries. Thus the polarization
expected from magnetic dust is at least 4.7\%. However, the CBI data
place a strict 3~$\sigma$ polarization upper limits of 4.8\% in the
specific intensity at the peak in $\rho$~Oph~W - and only 1\% in the
45~arcmin photometric aperture.

\subsection{Spinning dust} \label{sec:sd}

The spinning dust emissivities, being linear proportional to the
density $n_H$ (as dust emissivities in general), could be more
important in diffuse media relative to emission mechanisms based on
binary encounters, which are quadratic in $n_H$. With its very steep
low frequency spectral index, $\alpha \sim +3$, spinning dust explains
the lack of low-frequency emission.

Spinning dust gives a remarkably good fit to the $\rho$~Oph~W SED
(see~Fig.~\ref{fig:spec}) if $\beta = 1.7$, which is within the range
of observed values \citep[][]{dup03}. Higher emissivities result in a
drop at $\sim$60--90~GHz, which misses the WMAP~94~GHz data point.
The best fit spinning dust spectrum is obtained for the 'DC' case of
\citet{dl98b}.  Spinning dust requires that the product $ n_H f \sim
10^{5}$~cm$^{-3}$, where $f$ is the volume filling factor.

No local 31~GHz peaks are observed at the positions of S~1 and
SR~3. Although part of the emission surrounding S~1 and SR~3 could be
scattered light in IR reflection nebulae, the {\em Spitzer} IRS
area-spectroscopy highlights very bright PAH bands in the
circumstellar nebulae around S~1, SR~3 and $\rho$~Oph~W.  The absence
of S~1 and SR~3 in the CBI data may be inconsistent with spinning
dust, since the proximity of early type stars should result in high
spinning dust emissivities, provided VSGs are not depleted.  The
environment around S~1 corresponds to the 'RN' case of \citet{dl98b}:
the \citet{cas03} model atmospheres give a UV flux parameter $\chi
\approx 5000$ \citep[as defined in][]{dra96}. In this case the
predicted spinning dust emissivity per unit H-nucleus column density is
$10^{-17}~$~Jy~sr$^{-1}$~cm$^2$ at 31~GHz. The observed column
densities towards S~1 is $10^{23}$~cm$^{-2}$ (see
Sec.~\ref{sec:irdust}), so the emergent 31~GHz intensity from S~1
should be 1~MJy~sr$^{-1}$, which is a factor of 10 larger than the
observed peak 31~GHz intensities (which are found in $\rho$~Oph~W and
not toward S~1), and a factor of 40 larger than a strict upper limit
of 25~10$^{-3}$~MJy~sr$^{-1}$ in S~1 (see
Table~\ref{table:pahratios}).

We can take into account the possibility of VSG depletion by examining
the ratios of PAH and 31~GHz intensities. The dust-heating radiation
field can be parametrised, as in \citet{dl07}, by scaling the
interstellar radiation field in the solar neighbourhood \citep{mat83}.
The dimensionless parameter $U$ measures the average intensity of
radiation in 4$\pi$~sr, integrated from 0.09 to 8~$\mu$m. The spinning
dust emissivity per nucleon is remarkably independent of the intensity
of UV radiation \citep[][their Fig.~9]{dl98b}, while the flux in the
PAH bands is linear in $U$ \citep[for instance see Fig. 13
  in][]{dl07}. Therefore, if PAHs are involved in the 31~GHz emission,
the intensity ratio $R
=I_\nu(31\mathrm{GHz})/I_\mathrm{PAH}(11.3~\mu\mathrm{m})$ should be
inversely proportional to $U$. $R$ is independent of the PAH abundance
since it enters linearly both in $I_\nu(31\mathrm{GHz})$ and
$I_\mathrm{PAH}$.  Given that the spinning dust emissivities per
nucleon \citep{dl98b} vary by at most a factor of 10 for environments
with extremely different physical conditions, we can conservatively
assume that $R\times U$ should not vary by more than a factor of 10
when comparing widely different regions.

The IRS data allow extracting fluxes for the PAH band at 11.3~$\mu$m,
$F_\mathrm{PAH}(11.3~\mu$m), in two rectangular apertures near S~1,
one near SR~3, and one in $\rho$~Oph~W, indicated as black and
dotted-white boxes on Fig.~\ref{fig:zoom}. We subtracted a linear
continuum under the 11.3~$\mu$m band, as well as a sky background
PAH~11.3~$\mu$m intensity of $2~10^{-7}$~W~m$^{-2}$~sr$^{-1}$, obtained
by scaling the observed PAH intensities following the IRAC~8~$\mu$m
image. The background PAH intensity is only 5 times fainter than
observed in the S1~off aperture ($9.98~10^{-6}$~W~m$^{-2}$~sr$^{-1}$
before background correction).

Until we obtain higher-resolution data our estimates of the 31~GHz
specific intensities $I_\mathrm{31~GHz}$ are dependent on the MEM
model. In the brightest regions, as in $\rho$~Oph~W, we assume 10\%
uncertainties.  The product of the IRS extraction solid angle times
the CBI intensity at the centroid of the $\rho$~Oph~W aperture is an
approximation to $F_\nu(31~$GHz), the 31~GHz flux density.  For the
fainter regions we use the prior that the emission is due to spinning
dust (which is the aim of this test): The circumstellar nebulae around
S~1 and SR~3 are $\sim$3~arcmin in diameter and so should appear as
point sources in the CBI maps. A formal fit to the MEM visibility
residuals for point sources at the locations of S~1 and SR~3 gives
31~GHz flux densities of 1.3$^{+7}_{-4}$~mJy and $2.0^{+7}_{-3}$~mJy,
repectively, which can be converted into intensities assuming uniform
nebulae.  For S~1off we scale the 31~GHz intensity in the S~1 aperture
by the product of PAH~11.3~$\mu$m intensities times $U$ (i.e. as the
inverse squared projected distance).

The unattenuated intensity of dust-heating radiation, parametrised by
$U$, can be evaluated at the projected distance from the exciting
stars to the centre of each IRS aperture, using the model atmospheres
of \citet[][]{cas03}\footnote{ as given in {\tt
    http://wwwuser.oat.ts.astro.it/castelli/grids.html}} and the
stellar parameters given in Sec.~\ref{sec:stars}.  We obtain the
properties listed in Table~\ref{table:pahratios}. Taking into account
the uncertainties, the product $R\times U$ is consistent with
variations by less than a factor of 10. A marginal exception may be
the comparison between S~1off and $\rho$~Oph~W, where the 31~GHz
uncertainties are reduced relative to S~1 because of the hypothesis
that the 31~GHz intensity in S~1off scales with $U \times
I_\mathrm{PAH}$ relative to S~1. In the S~1off - $\rho$~Oph~W
comparison, taking 3~$\sigma$ shifts in $R\times U$ can bring its
variations down to a factor of 4, and so in agreement with spinning
dust.

A source of uncertainty is the correction for UV attenuation, which
may affect all lines of sight in
Table~\ref{table:pahratios}. Attenuation is most important for S1~off,
since the circumstellar nebula around S~1 is probably
ionisation-bounded (see Sec.~\ref{sec:irdust}). The value in
Table~\ref{table:pahratios} is therefore an upper limit. Similarly for
SR~3, which is obscured in the visible. On the other hand S~1 is
bright in the visible, so its circumstellar nebula stems from the
walls of a wind-blown cavity driven by S~1 into the $\rho$~Oph~A
core. Thus UV attenuation should be negligible in the S~1 IRS
aperture. According to the model of \cite{lis99}, HD~147889 is
separated from LDN~1688, so that its UV luminosity should be fairly
unatenuated towards $\rho$~Oph~W.

\begin{table}  % see ~/ROPH/README
\centering
\caption{Observations derived from the CBI and PAH~11.3~$\mu$m
  intensities in the {\em Spitzer} IRS apertures.}
\label{table:pahratios}
\begin{tabular}{lrrrrr}
\hline 
                  & $R \times U^{a,b}$    &    $ U^c $   & $I_{31\mathrm{GHz}}^d$  & $I_\mathrm{PAH}^e$   &  $d^f$   \\ 
 $\rho$~Oph~W     &  27.5$\pm$2.7          &   66           &  2.2$\pm$0.2(-1)       &   3.3(-6)       &  10.4  \\
       SR~3        &  $<$1.2$\pm$3.2          &   $<$238    &  3.2$\pm$8.4(-3)       &   3.8(-6)       &  0.97  \\
        S1         &  1.0$\pm$4.9          &   1076         &  1.7$\pm$8.4(-3)       &   1.1(-5)       &  0.92  \\
        S1 off        &  $<$1.0$\pm$1.1          &   246    &  5.2$\pm$5.9(-4)       &   8.0(-7)       &  1.92  \\
\hline
%                  & $R \times U^{a,b}$    &    $ U^c $   & $I_{31\mathrm{GHz}}^d$  & $I_\mathrm{PAH}^e$   &  $d^f$   \\ 
%$ \rho$~Oph~W     &  12.0         &   40.08      & 1.93e-01  2.17e-01    &  3.45e-06        &  10.4  \\ % 55942.0289855072 
%      SR~3        &  39.5         &   238.1      & 1.25e-01  1.52e-01    &  4.03e-06        &  0.97 \\  % 31017.3697270471 
%       S1         &  22.9         &   1076       & 4.61e-02  8.25e-02    &  1.16e-05        &  0.92 \\  % 3974.13793103448 
%   S1 off         &     1         &   245.7      & 7.59e-04  4.91e-02    &  9.98e-07        &  1.92 \\  % 760.521042084168 
%$ \rho$~Oph~W     &  38.36         &   128.13    & 1.93\,10$^{-1}$    &  3.45\,$10^{-6}$        &  10.4  \\ % 55942.0289855072 
%      SR~3        &  53.17         &   320.31    & 1.25\,10$^{-1}$    &  4.03\,$10^{-6}$        &  0.97 \\  % 31017.3697270471 
%       S~1         &  22.89         &   1076.2    & 4.61\,10$^{-2}$    &  1.16\,$10^{-5}$        &  0.92 \\  % 3974.13793103448 
%   S~1 off         &      1         &   245.68    & 7.59\,10$^{-4}$    &  9.98\,$10^{-7}$        &  1.92 \\  \hline % 760.521042084168 
\end{tabular}
\begin{flushleft}
$^a$  $R=I_\nu(31\mathrm{GHz})/I_\mathrm{PAH}(11.3~\mu\mathrm{m})$
$^b$  Normalized to the value in the S~1~off aperture. 
$^c$  Unattenuated UV field parameter. 
$^d$  CBI MEM specific intensities in MJy~sr$^{-1}$.
$^e$  PAH~11.3$\mu$m intensity in W~m$^{-2}$~sr$^{-1}$. 
$^f$  Projected distance from exciting star, in arcmin.
$^g$  parenthesis indicate a  power of ten exponent multiplying both
  the average values and their uncertainties. 
\end{flushleft}
\end{table}

Could physical conditions in S~1 and SR~3 be such that spinning dust
is quenched by mechanisms not contemplated by \citet{dl98b}?  The
possibility of vanishing grain polarization can be discarded.
According to the PAH ionisation diagnostic of \citet[][ their
  Fig.~16]{dl01}, the PAH band fluxes given by \citet{bou96},
extracted from an aperture 10~arcmin north of $\rho$~Oph~W, correspond
to neutral PAHs. Any photoelectric charge expected in the intense
UV-fields of S~1 and SR~3 increases quadratically the spinning dust
emissivity \citep[Eq.~11 in][]{dl98b}. But it could be that the
radiation field from HD~147889 is sufficiently hard to boost the
spinning dust emissivity through VSG-ionisation, and yet not
completely deplete the grains.  We used the IRS spectra to place S~1
in the PAH-colour diagram of \citet[][their Fig.~16]{dl01}, finding it
lies close to $\rho$~Oph~W - i.e. at the locus of neutral
PAHs. Another interesting possibility is that perhaps PAH rotation
brakes to lower frequencies in the vicinity of SR~3 and S~1, maybe as
a consequence of grain alignment in intense magnetic fields
\citep[which are not considered in][]{dl98b}.

Yet another alternative to explain the lack of 31~GHz emission from
S~1 and SR~3 is that the spinning VSGs are not related to PAHs, so
that they do not emit in the PAH bands.  For instance they could
perhaps be nano-silicates \citep[e.g.][]{wit98} very fragile to UV
radiation, or maybe chain-like carbonaceous molecules.  If so the
existing spinning dust models must be revised, since they are
calculated for aromatic carbonaceous grains, including a population of
sheet-like grains (i.e.  PAHs).

\citet{igl06, igl05} has proposed hydrogenated fullerenes as a carrier
for the 31~GHz emission. Perhaps the polarized fulleranes could be
depleted near S~1 and SR~3.  But at present we lack diagnostics of such
fulleranes to test this alternative. As proposed by \citet{web92}
fulleranes could also be the carriers of the diffuse interstellar
bands (DIBs). Spectroscopy of stars in the background of $\rho$~Oph~W
is required to examine the possibility of a link between the DIBs and
the anomalous foreground.

\section{C\,{\sc i} continuum} \label{sec:CI}

\subsection{Cold plasmas in PDRs}

The outer layers of molecular clouds are composed of atomic
hydrogen. The ionisation energy of H$_2$ is 15.4~eV, while the H$_2$
dissociation continuum starts at 14.7~eV, both of which are attenuated
by H\,{\sc i} ionisation.  In practice H$_2$ photodissociation occurs
in the radiative cascade following bound-bound absorption of photons
with discrete energies in the range 11.3 -- 13.6~eV
\citep[e.g.][]{tie05}. Thus the bulk of the $\rho$~Oph cloud is
largely neutral.

All photons harder than 13.6~eV are absorbed in the vicinity of
HD147889, producing a $\sim 5$~arcmin diffuse H\,{\sc ii} region
(about 0.2~pc in physical size). Yet a residual charge density exists
in the neutral layers of $\rho$~Oph exposed to radiation from
HD147889.  C-ionising photons with energies in a continuuum between
11.3~eV and 13.6~eV penetrate deep inside the exposed neutral layers
of $\rho$~Oph~W. C-ionisation extends into molecular layers. Other
metals with ionisation potentials inferior to 13.6~eV, most notably
sulfur, also contribute to the ionisation fraction $x = n_e / n_\mathrm{H}$,
but in proportion to their abundance.

Another source of ionisation in molecular clouds is the cosmic-ray
ionisation of H$_2$. For an isolated molecular cloud subjected to the
average interstellar cosmic ray ionization rate $\zeta$, molecular
line observations have yielded $x \approx 10^{-8}$. Specifically for
$\rho$~Oph \citet{woo79} give $x =2.6~10^{-7}$. However the molecular
line diagnostics in \citet{woo78} trace the dense core in
$\rho$~Oph~A, and may not be representative of $\rho$~Oph~W. Is it
possible that in $\rho$~Oph~W the vicinity of HD147889 increases
$\zeta$ sufficiently so that H$_2^+$ also contributes to $x$? Probably
not, since the ratio of projected distances between $\rho$~Oph~A and
$\rho$~Oph~W to HD147889 is $\sim$2, which would only raise $\zeta$ by
a factor of 4, and yield an $x$ value still 2 orders of magnitude
below the carbon abundance.

%We have no data on $\zeta$ in $\rho$~Oph~W, and so cannot incorporate
%H$_2^+$ in the calculations below.

\subsection{C\,{\sc i} model}  \label{sec:CImodel}

Here we estimate the expected intensity levels of free-free emission
due to C$^+$ encounters with free electrons (C\,{\sc i} continuum)
from the cold plasma in $\rho$~Oph~W.  Although the ionisation
fraction in PDRs is very small, $x \sim 10^{-4}$, the free-free
emissivity is $\propto 1/\sqrt{T_e} $ and biased towards lower
temperatures. A temperature of $\sim$50~K is representative of most of
the emissivity-weighted C\,{\sc ii} region in the models of
\citet{lis99} and \citet{hab03}. A shortcoming of the present model is
that it requires a clumpy medium with very high H-nucleus densities, of
order $10^7$~cm$^{-3}$, and small volume filling factors, $f \sim
10^{-4}$. 

% see clump_size_BB.pl

The available data lead us to consider an ensemble of dense neutral
clumps, with an ionisation fraction $x \sim 10^{-4}$ (the abundance of
C), a total extent $\theta_N \approx 20$~arcmin, and covering a solid
angle $\Omega_N \approx \pi (\theta_N /2)^2$. Since $\rho$~Oph~W is
undetected in the HRAO data of \citet{baa80}, with a 20~arcmin beam
and a rms sensitivity of 30~mK, at 2.3~GHz the total emission from the
clumps should be less than $ F_c(2.3~\mathrm{GHz}) \approx \Omega_N
\times 3 I_\mathrm{rms}(2.3~\mathrm{GHz}) \approx 2~$Jy. The total
flux density seen by {\em WMAP} at 33~GHz and not attributed to
H\,{\sc i} free-free is 12.3~Jy. Thus the clumps must be optically thick
at 2.3~GHz, and cover a solid angle inferior to $\Omega_c \approx
F_c(2.3~\mathrm{GHz}) / B_\nu(2.3~\mathrm{GHz}, T_e = 20~ \mathrm{K})
\approx \pi~ 1.5^2$~arcmin$^2$, where $B_\nu$ is the Planck
function. Additionally, the clumps, while still optically thick at
5~GHz, are not detected in PMN, with a beam $\theta_\mathrm{PMN} =
3.7$~arcmin FWHM and an rms noise of
$I_\mathrm{PMN}\sim$100~mJy~beam$^{-1}$ =
0.67~MJy~sr$^{-1}$\footnote{in which we have taken into account
  correlated pixels with a factor $\sqrt{N_\mathrm{beam}}$, where
  $N_\mathrm{beam}$ is the number of pixels that fall in one
  3.7~arcmin beam}. Therefore an individual clump must be beam-diluted
in PMN and cover a solid angle less than $(I_\mathrm{PMN} /
B_\nu(5\,\mathrm{GHz}, 20\,\mathrm{K}) ) (\pi/(4 \ln(2)))
\theta_\mathrm{PMN}^2 \approx 3~\mathrm{arcmin}^2$.

As discussed in Sec.~\ref{sec:spec}, the CBI sees about
30--40\%\footnote{this is the ratio of 31~GHz flux densities obtained
  without the use of a prior to the {\em WMAP}~33~GHz flux densities}
of the total 31~GHz emission.  The clump distribution should be
sufficiently dense so as to mimick a uniform signal at CBI
resolutions, while also extending to $\theta_N$. Thus the inter-clump
separation should be (a) less than 8~arcmin, the highest spatial
frequency in the CBI $uv$-plane for the configuration used here, and
(b) greater or similar to the PMN beam.  To make headway, we will
assume a number of clumps $N_c = (\theta_N / \theta_\mathrm{PMN})^2
\approx 30$.

A plane-parallel geometry seen edge-on gives the highest clump
opacities. To maximize the cold free-free opacity given an electron
density, each clump would have to be $\theta_\mathrm{PMN}$ in length,
equally deep, and only $\Omega_c / (N_c \theta_\mathrm{PMN} ) \approx
4~$arcsec wide. Higher densities alleviate the need for such large
aspect ratios.

What is the density required to explain the 31~GHz flux density with a
cold plasma, if we ignore the lower frequency data? In a uniform disk
nebula 10~arcmin in diameter, and equally deep, with a C\,{\sc ii}
region temperature of 150~K, we find that electron densities of $n_e =
$60~cm$^{-3}$ reach the observed 31~GHz level. The same result is
arrived at by fixing the opacity profile to that of the CBI MEM
model. Thus H-nucleus densities of only $n_H \sim 10^5-10^6$~cm$^{-3}$
(within the range observed in $\rho$~Oph~W, see
Sec.~\ref{sec:physconds}) account for all of the CBI flux.

\subsection{SED fit} \label{sec:CI_model}

%perldl> $Vclumps = 30. * (2.*3.7*3.7);
%perldl> $Vneb = ((10.*60)**2 * 3.14 * 20); 
%perldl> p ($Vclumps/$Vneb);
%3.63322717622081e-05

The C\,{\sc i} continuum model shown in Fig.~\ref{fig:spec}~b consists
of a clump ensemble with $N_c = 30$, an electron density of
750~cm$^{-3}$, and $T_e= 20$~K. Each clump is a narrow sheet only
2~arcsec wide\footnote{We chose 2 instead of 4~arcsec to better
  accomodate the 2.3~GHz point}, 3.7~arcmin long, and equally
deep. The corresponding volume filling factor is $f \sim
10^{-4}$. Uniform disk clumps, with equal depth and diameter and
subtending the same solid angle as the sheets, would require electron
densities of 2200~cm$^{-3}$. We favour a sheet-like geometry because
of the considerations in Sec.~\ref{sec:stars}: The thin sheet
geometries could result from limb-brightening of a C\,{\sc ii} shell
centred on HD147889.

The total H-nucleus densities required by the C\,{\sc i} model,
whatever the geometry of the clumps, are of order $\sim
10^7$~cm$^{-3}$, if all of C is ionised. The corresponding C\,{\sc ii}
region mass is 153~M$_\odot$ for sheets, or 45~M$_\odot$ for disks,
both of which are consistently less than the total mass enclosed by
regions with finite $I_{31~\mathrm{GHz}}$ (see Sec.~\ref{sec:irdust}).

In order to further test the C\,{\sc i} continuum model we searched
the ATCA archive for pointings near $\rho$~Oph~W.  $\rho$~Oph~10
\citep[also known as Doar 21;][]{fal81} was observed in aray
configuration C397, at frequencies of 1376, 2378, 4800 and
8640~MHz\footnote{ATCA programme C397, PI E. Feigelson}. The phase
centre lies in the 31~GHz $\rho$~Oph~W ridge, 2.2~arcmin south of the
peak. At 4.9~GHz the ATCA primary beam is 5~arcmin. Yet no clumpy
medium is detected with a noise of 5mJy/beam. Clearly the putative
C~\,{\sc i} clumps, if they exist, should be beam-dilluted even inside
the 12$\times$7~arcsec$^2$ beam: the optically thick flux density in
one synthetic beam is 160~mJy, if $T_e = 20~$K. Very thin sheets,
$\sim$~1~arcsec wide, could still fit in.

\section{Radio recombination lines} \label{sec:RRLs}

Radio recombination lines (RRLs) are concomitant to free-free
continua.  Carbon RRLs have indeed been observed towards $\rho$~Oph,
albeit at low resolutions \citep[][]{bro74a}. Here we provide upper
limits on the RRL system in $\rho$~Oph~W, which require rather cold
C\,{\sc ii} regions ($T_e \sim 20-50$~K).

\subsection{RRL data from the literature}

\citet{pan78} examined the most complete set of RRL data towards
$\rho$~Oph to date. They mapped the neighbourhood of S~1, but did not
extend their coverage to $\rho$~Oph~W, unfortunately. The highest
frequency RRLs considered by \citet{pan78} are C90$\alpha$ and
C91$\alpha$, at $\sim$9~GHz, which they interpreted as stemming from
circumstellar gas about S~1, with electron densities $n_e \sim
15~$cm$^{-3}$ and $T_e \sim 150~$K. This circumstellar C\,{\sc ii}
region was inferred to be less than $\sim$2~arcmin in diameter, and
surrounded by a diffuse halo with $n_e \sim 1~$cm$^{-3}$, traced by
the lower frequency carbon RRLs. The continuum level expected from
such a C\,{\sc ii} region is $\sim$1~mJy at 31~GHz, consistent with
its absence from the CBI MEM model.

\subsection{CBI limits on Ka-band RRLs}

The brightest carbon RRLs expected in the 10 CBI channels are
C57$\alpha$ to C62$\alpha$. Four 1~GHz wide channels, centred at 27.5,
30.5, 33.5, and 35.5, are free from $\alpha$ RRLs. Yet the CBI flux
densities in each channel, inferred by cross-correlation with a
processed IRAC~8$\mu$m template, are scattered about a single power
law; no decrement is seen in the four channels devoid of RRLs.  A
1~$\sigma$ upper limit to the contribution of RRLs in a single channel
is given by the rms dispersion about the best fit power-law, of
$\sigma_\mathrm{RRL} = 0.04$~Jy. 

The expected RRL flux in the CBI channels is quite close to the above
upper limit. In the C\,{\sc ii} region model of Sec~\ref{sec:CI}, the
flux density from C62$\alpha$, at $26.95~$GHz, in the 1~GHz wide CBI
channel centred on 26.5~GHz, is 0.6~Jy under LTE conditions. However,
for $n < 100$ the excited energy levels leading to the emission of
$\alpha$ photons are severely depopulated relative to LTE at
$\sim$150~K \citep[][]{sal79, wal82}\footnote{We have cross-checked
  our calculations against those in \citet{pan78} and Eq.~3 in
  \citet{hei96}}. The exact level population in $n = 62$ is very
sensitive on temperature and dielectronic recombination. The value
given by \citet[][their Fig.~3]{wal82} is $b = 0.5$, relative to LTE,
implying a C62$\alpha$ contribution of 0.3~Jy, which should be
detectable by the CBI\footnote{We have neglected the effects of
  stimulated emission, accounted for in the $\beta$ coefficient of
  \citet[][their Fig.~5]{wal82}, because $\beta \approx 1$ for the
  transitions considered here}.  Lowering $T_e$ to 50~K brings the
predicted RRL flux down to 3~$\sigma_\mathrm{RRL}$, i.e. marginally
consistent with the CBI upper limit.
     
\subsection{Mopra limits on K- and W-band RRLs}

The detection of the RRL system of $\rho$~Oph~W in the C\,{\sc i}
model of Sec~\ref{sec:CI_model} would be very difficult. The ensemble
of clumps is optically thick below $\sim$10~GHz.  At higher
frequencies the excited energy levels leading to the emission of
$\alpha$ photons are severely depopulated relative to LTE
\citep[][]{sal79, wal82}.

We nonetheless attempted to detect the RRL system of $\rho$~Oph~W
using the UNSW-MOPS spectrometer (Sec.~\ref{sec:mops}), and found no
RRLs at the expected radial velocity, $V_\mathrm{lsr} =
+3~$km~s$^{-1}$ \citep{bro74a,pan78}. With the line profiles observed
at lower frequencies, of 1.5~km~s$^{-1}$ FWHM, we can estimate the
expected contribution of RRLs in the Mopra datacubes and place upper
limits on the high frequency RRLs.

The ensemble of C\,{\sc ii} clumps subtends a solid angle of less than
$\Omega_c = \pi 1.5^2$~arcmin$^2$. If there are $N_c = 30$ clumps
spread uniformly over 20$\times$20~arcmin$^2$, then only one clump is
expected on average in each 5~arcmin pixel. The expected intensity
from one clump, covering about $\pi ~8^2$~arcsec$^2$ is diluted in
5$\times$5~arcmin$^2$. At $T_e = 150~$K, the emergent peak LTE
intensity in C73$\alpha$ from one clump is 1262~MJy~sr$^{-1}$, or
2.97~MJy~sr$^{-1}$ when diluted in the 5~arcmin pixels. In dense and
cold gas dielectronic recombination enhances the LTE departure
coefficients to $b \sim 1$. Thus with a 3~$\sigma$ upper limit of
1.1~MJy~sr$^{-1}$ on C73$\alpha$, the Mopra data rule out `warm'
temperatures (note that collisionally excited lines, such as [C\,{\sc
    ii}]~158$\mu$m, are biased towards higher temperatures). By
contrast, at $T_e = 20~$K, the peak LTE RRL intensity in 5~arcmin
pixels is 0.38~MJy~sr$^{-1}$, and $b \sim 0.3$, so that the expected
C73$\alpha$ intensity is only 0.12~MJy~sr$^{-1}$ - consistent with its
non-detection.

% !!!!!!!!!!!!!!!!!!!!!!!! give exact b value . 1982ApJ...260..317W.pdf 

%For C73$\alpha$, the Mopra 3~$\sigma$ upper limit is
%1.1~MJy~sr$^{-1}$, averaged over 5~arcmin pixels, which rules out
%20$\times$20~arcmin$^2$ uniform slab models for the 31~GHz flux
%density.
%
In the W band the $b$ coefficients are vanishingly small at
20~Ktemperatures. For C42$\alpha$ and $T_e = 20~$K, we approximate to
the $n=50$ case of \citet{sal79} (for $\log(N_e~ /~ 1~\mathrm{cm}^{-3}
)=2.5$ and with a 100~K background): $b = 5~10^{-4}$. The expected
C42$\alpha$ intensity in the 5~arcmin pixels is $b\times
9.6$~MJy~sr$^{-1}$, while the rms noise is 188~MJy~sr$^{-1}$. 

\section{Discussion}   \label{sec:disc}

The [C\,{\sc ii}] $\lambda$158~$\mu$m image of $\rho$~Oph~A by
\citet{yui93}, with a 15~arcmin beam, shows that C$^+$ is present and
that [C\,{\sc ii}] $\lambda$158~$\mu$m shares a similar extent as the
cm-wave continuum. It is remarkable that S~1 is not coincident with a
[C\,{\sc ii}] $\lambda$158~$\mu$m peak.  In this Section we discuss
the likely emission levels of C\,{\sc i} continuum in the $\rho$~Oph
environment, and consider the possibility of combining both C\,{\sc i}
and spinning dust.

\subsection{Exciting  stars}  \label{sec:stars}

%, or $T_\mathrm{eff} \sim
%20300~$K, $\log (g/\mathrm{cm~s}^{-2} ) = 3.4$

\subsubsection{HD147889}

Atmospheric parametres for the two components of HD147889 were
obtained by fitting ``Tlusty'' non-LTE model atmospheres \citep{lan07}
to the FEROS spectra (Sec.~\ref{sec:feros}), with rotation velocities
$v\sin i$ of 50 and 30~km~s$^{-1}$ respectively, and a surface area
ratio of 1.3. The temperatures are 23000 and 20000~K and the surface
gravity $\log( g) $ is 4.25 for both components ($\log(g) < 4.0$ or
$>4.5$ are ruled out). The surface gravity is determined mainly from
the strength of the line wings of H$\gamma~\lambda 4340$. However,
this strength also depends on the temperature. The temperature mainly
comes from the strength of He\,{\sc i} $\lambda 4387$ and $\lambda
4471$ as compared with Mg\,{\sc ii} $\lambda 4481$ and the Si\,{\sc
  iii} triplet $\lambda\lambda\lambda$ 4552, 4567, 4573.

The heliocentric velocities of each component were +64,
$-$99~km~s$^{-1}$ on UTC = 2006-02-07, $-52$~km~s$^{-1}$,
$+53$~km~s$^{-1}$ on UTC = 2008-05-16, and $-39$~km~s$^{-1}$,
$+33$~km~s$^{-1}$ on UTC = 2008-05-17. This gives us a systemic
heliocentric velocity of $-7~$km~s$^{-1}$, or a local standard of rest
velocity of $+3~$km~s$^{-1}$ (so equal to that of the carbon RRLs),
and a mass ratio of 1.3.

If the two components of HD147889 are on the main sequence, as
indicated by their surface gravities, then the stellar radii and
masses can be inferred by interpolating the zero-age main-sequence
models of \citet{sch92}. We obtain radii of 3.52 and 3.10~R$_\odot$,
and masses of 8.31 and 6.36~M$_\odot$. But it must be borne in mind
that the IR excess of HD147889 and its interaction with the $\rho$~Oph
star forming cloud make it a likely pre-main sequence star, in the
sense that it is still contracting towards the zero-age main
sequence. The models of \citet[][their Table~1]{beh01} show that at an
age of $2.4~10^5$~yr (so close to $10^6$~yr, the approximate lifetime
of dark clouds), a 12~M$_\odot$ star would still be contracting, with
a surface temperature of $22~000$~K and a radius of 10.3~R$_\odot$.
To make headway the spectral types we adopt are B2IV and B3IV, but we
note that we lack a precise measurement of the stellar luminosities.

%%%
% (compared to
%**** on the ZAMS) quote ZAMS value for Radius and Luminosity. 

The line of sight to HD147889 appears to be strongly affected by
extinction from LDN~1688. The H-nucleus column density map reported in
Sec.~\ref{sec:irdust} gives $N_H = 1.13~10^{22}$~cm$^{-2}$ in the
direction of HD147889, or $A_\mathrm{V} = 6.76$ using the conversion
factors from \citet{dra03} with $R_\mathrm{V} = 3.1$. The
$A_\mathrm{V}$ map from \citet[][]{rid06} gives $A_\mathrm{V} = 8.1$
at the position of HD147889. However, a comparison between the
observed $B - V$ colour and that expected from the model atmospheres
from \citet{cas03} gives $E(B-V) = 0.8$ (or $A_\mathrm{V} \sim 2.5$).
Yet the stellar radii inferred above require extinction values
$A_\mathrm{V} = 4.5$ to be compatible with the photometry. Hence the
extinction towards HD~147889 is manifestly very structured on small
scales, and requires an exceptionally large $R_\mathrm{V} = 5.6$ or
higher if HD~147889 is still contracting.

%This low extinction value implies unrealistically small stellar radii,
%so either the extinction law towards HD147889 or its spectrum are
%peculiar.

%Adopting $A_\mathrm{V} = 6.76$ gives a radius of 10.4~R$_\odot$ for
%HD147889A, and 8.5~R$_\odot$ for HD147889B, in accordance with
%luminosity class IV/III.
%

%I just tried to fit the Feros spectrum with the Tlusty models published 
%by Lanz, T., & Hubeny, I. 2007, ApJS, 169, 83. I attach a plot with my 
%current best fit. The model is plotted in red. The parameters are as 
%follows:
%
%radial velocities: +55, -107
%temperatures: 22000, 19000
%log g: 4.0, 4.0
%v sin i: 50, 30 km/sec
%flux ratio: 1.5
%
%The surface gravity is determined mainly from the strength of the line 
%wings of Hgamma (4340). However, this strength also depends on the 
%temperature. The temperature mainly comes from the strength of HeI 4387 
%and 4471 as compared with MgII4481 and the  SiIII triplet 4552, 4567, 4573.
%

\subsubsection{S~1 and SR~3}

S~1 is a close binary system composed of a B4V star \citep[][]{lad88}
and a K-type companion with a 10~mJy peak at 10~GHz \citep[including
  its 2~mJy radio halo,][]{and88}. For S~1, $T_\mathrm{eff} \sim
15800~$K, and surface gravity $\log (g /\mathrm{cm~s}^{-2}) = 4.0$.
The spectral type we adopt for SR~3 is B6V \citep{eli78}, or
$T_\mathrm{eff} \sim 14000~$K, $\log (g/\mathrm{cm~s}^{-2} ) =
4.0$. \citet[][their Sec.~IIIb]{lad88} and \citet{bon01} give
bolometric luminosities $L_\star$ for S~1 and SR~3. For S~1, $L_\star
= 1100~L_\odot$, which is consistent with B4V, but for SR~3, $L_\star
= 100~L_\odot$, which is too low for B6V. The spectral type of SR~3 is
probably closer to B9V.

%#cloudy models, hden6 
%r_S~1 0.0825257552137707 Omega_S~1 0.0213958176268231 
%r_SR~3 0.0651953466188788  Omega_SR~3 0.0133531297809003 
%r_W 0.0264082416684066 Omega_W 0.105632966673626
%

\subsection{Physical conditions}  \label{sec:physconds}

%They require a clumpy PDR with a typical source size of
%$\theta_\mathrm{c}>\sim 30''$, and area filling factor $\sim0.16$, or
%$f\sim 0.06$.

The 1.1~mm continuum data in \citet{you06} are strongly high-pass
filtered and trace the smaller angular scales. They find an average
density of $\sim 10^{6}~$cm$^{-3}$ in 44 molecular cores containing
$\sim$80~M$_\odot$, or about 1/20--1/100 the mass of the entire
$\rho$~Oph complex (see Sec.~\ref{sec:irdust}). 

A density of $n_H = 10^4 - 10^5$ is given by \citet{lis99} from PDR
models of the far-IR line ratios. They fit [O\,{\sc i}] 63,
145~$\mu$m, while the [C\,{\sc ii}]~158$\mu$m levels can only be
reproduced with $n_H=10^6$~cm$^{-3}$ and an unrealistically low field
$G_\circ$. Higher densities than $10^6$cm$^{-3}$ are required to
accomodate $G_\circ \sim 100$ inferred by \citet[][]{lis99}. They
treated HD147889 as a main sequence star, and used a distance of
150~pc. With the Hipparcos distance to HD147889 of 135~pc and
luminosity class IV we obtain $G_\circ = 629$, using the model
atmospheres from \citet[][]{cas03}.

Bypassing the PDR models, we can infer a H-nucleus density from the
[C\,{\sc ii}]~158$\mu$m cooling rate \citep[e.g.][his
  Eq. 2.67]{tie05}. The fluxes from \citet{lis99} then give $n_H \sim
3~10^{4} - 2~10^{6}~$cm$^{-3}$ for temperatures of $10-100~$K and a
filling factor $f = 10^{-3}$. Note, however, that the critical density
of [C\,{\sc ii}]~158$\mu$m is 2.7~10$^{3}$~cm$^{-3}$ at 100~K, so that
it does not trace the dense clumps required by the C\,{\sc i}
continuum model.

\subsection{C\,{\sc i} Str\"omgren spheres}  \label{sec:stromgren}

%A prediction of the cold plasma model is that the $\rho$~Oph~W region
%should be seen in absorption against the diffuse warm-ionised medium
%at 2-5~GHz.

%We use the hydrogenic total recombination coefficients $\alpha_B$ from
%\citet{sto95}, in case B of \citet{bak38}. 
% in case B of \citet{bak38}.

The model atmospheres from \citet{cas03} predict that the C-ionizing
luminosity of HD147889 is $\sim$25 times that of S~1 and 612 times
that of SR~3.  From its vantage point at the back of the $\rho$~Oph
main cloud HD147889 illuminates the entire cloud \citep{lis99}. By
contrast S~1 is surrounded by its disk-like circumstellar nebula
(which probably stems from the clumpy walls of a wind-blown cavity,
see Sec.~\ref{sec:irdust}). Since S~1 is optically visible a fraction
of its UV luminosity escapes. The small part of it's carbon-ionising
flux that is absorbed in the dense circumstellar disk radiates a
meagre 2~mJy at 5~GHz \citep{and88}.

If the C\,{\sc ii} regions of $\rho$~Oph are ionisation-bounded
nebulae their average electron density is given by ionisation balance:
$n_e^2= S / \alpha V$, where $V$ is nebular volume and $S$ is the
stellar luminosity in C-ionising photons (below H-ionising energies).
We use the total recombination coefficients $\alpha$ from
\citet{nah97}, assuming $T_e = 100~$K.  For optically thin radiation,
the total C\,{\sc ii} region flux density is 
\begin{equation}
F_\nu = S B_\nu \kappa^1_\nu / \alpha, \label{eq:CIIstromgren}
\end{equation}
where $B_\nu$ is the Planck function, and $\kappa^1_\nu$ is the
C\,{\sc i} free-free opacity for unit electron density.

For S~1, we obtain $S = 9.9~10^{45}$~s$^{-1}$, which for a closed
geometry should give an integrated flux density of 458~mJy at 31~GHz.
S~1 is not detected in the CBI maps. In Sec.~\ref{sec:sd} we report a
31~GHz point-source flux density of 1.3$^{+7}_{-4}$~mJy for S~1. It appears
that the circumstellar nebula around S~1 must be disk-like, or else
sufficiently clumpy that $\sim$90~\% of C-ionising photons escape the
nebula. This is consistent with the fact that S~1 is an optically
visible star (by contrast with SR~3).

For SR~3 we have $S = 4.1~10^{44}$~s$^{-1}$. The predicted flux
density in a closed geometry is 20~mJy at 31~GHz, consistent with its
non-detection by the CBI (the 31~GHz point-source flux density of SR~3
is $2.0^{+7}_{-3}$~mJy).

If HD~147889 has settled on the main sequence, we find $S =
1.4~10^{47}~$s$^{-1}$, and an upper limit for the 33~GHz flux density
of 6.5~Jy.  The background-subtracted 33~GHz flux density inside the
CBI PB aperture of 45~arcmin is 6.7~Jy, of which $\sim$1.5~Jy stem
from the diffuse H\,{\sc ii} region surrounding HD~147889. This is
close to the levels required for an ionisation bounded and spherical
nebula. But part of the C-ionising radiation should escape the
nebula. Thus for the C\,{\sc i} continuum interpretation it is
necessary that HD~147889 be a pre-main sequence binary star that is
still contracting, with a primary stellar radius of about
4.5~R$_\odot$, and a mass of $\sim$10~M$_\odot$, so that $\log(g) \sim
4$ to 4.5 (as inferred from the FEROS spectroscopy), and a secondary
mass of 7.7~M$_\odot$.

The C\,{\sc ii} Stromgr\"en sphere around HD~147889 extends out to
$\sim$10~arcmin, the projected distance to $\rho$~Oph~W. For a filled
sphere, the required electron density is $\sim$100~cm$^{-3}$, implying
H-nucleus densities of $\sim10^6$cm$^{-3}$.  Such a C\,{\sc ii} region
would be optically thin at 5~GHz and reach $\sim$10~Jy, which is ruled
out by the data. Thus the C\,{\sc ii} region around HD~147889 must be
a thin shell, as indicated by the geometry of $\rho$~Oph~W. However,
in order to reach the densities required by the SED model of
Fig.~\ref{fig:spec}, the thickness of the $\rho$~Oph~W C\,{\sc ii}
region should be only 2--3~arcsec, which is surprisingly thin. Such a
thin shell could be replaced by an ensemble of flat clumps seen
edge-on (see Sec.~\ref{sec:CI_model}).

We have modelled the C\,{\sc ii} regions in $\rho$~Oph using the
Cloudy photoionisation package \citep[version c07.02.01, last
  described by][]{fer98}. We treat $\rho$~Oph~W as a shell in an open
geometry, separated from HD147889 by its projected distance, while S~1
and SR~3 are embedded in ionisation-bounded circumstellar nebulae,
with a fiducial inner radius of $10^{15}$~cm. The predicted C\,{\sc
  ii} layer in $\rho$~Oph~W is extremely narrow, with a width $\delta
<$1.2~arcsec at densities $>10^{6}$~cm$^{-3}$, which is consistent
with the thin ridge model. The solid angle subtended by the C\,{\sc
  ii} region in $\rho$~Oph~W, which consists of an ensemble of $N_c =
30$ sheets, each 3.7~arcmin long, is about $30 \times 3.7 \times
\delta / 60$~arcmin$^2$. This is a factor of $\sim$100 larger than in
the C\,{\sc ii} regions around S~1 and SR~3 (if the H-nucleus density
is constant at 10$^{6}$~cm$^{-3}$).

%The RRL system concomitant to the diffuse C\,{\sc i} sources may be
%detectable. For a diffuse cold plasma with $T_e = 150~$K, $N_e = 21~
%$cm$^{-3}$ (Sec.~\ref{sec:CI}), example RRLs intensities are 

%For a carbon abundance of [C/H]$ = -4$~dex, the C\,{\sc ii} region
%mass is about $(\theta_N D)^3 f \langle n_e \rangle 10^\mathrm{[C/H]}
%= 65~$M$_\odot$, for a nebular diameter $\theta_N = 20$~arcmin.
%

%COMMENT 2MASS flux. 
%I_J = 0.1 erg/s/cm-2/Hz/sr  (surface brightness)
%I_H = 0.06 I_J
%I_K = 0.025 I_J 
%Padoan P., Juvela M., Pelkonen V.-M., 2006, ApJ, 636, 101

\subsection{Origin of the 2MASS diffuse emission}

The 2MASS~K$_s$ band emission from $\rho$~Oph~W is coincident with the
31~GHz ridge.  What is the nature of the nebulosity seen in the 2MASS
J,H, and K$_s$ images?  All three 2MASS bands share a similar morphology
with the IRAC images, although not exactly coincident. 

\subsubsection{Problems with near-IR scattered light, VSG continuum,  and   free-free}

We estimated that the diffuse intensities in $\rho$~Oph~W drop at
shorter wavelengths, in ratios J:H:K $\sim 1:2.2:2.7$\footnote{we used
  the zero points for the Montage mosaics, took the median intensity
  values in a square box 20 pixels on a side, centred on
  J2000~16:25:57.88~--24:21:11.7, and subtracted a median background
  extracted from a similar box but centred on
  J2000~16:25:58.10~--24:14:42.4}. This is at odds with the colours
expected from scattered light. \citet{pao06} calculate that the
emergent intensities of scattered light in neutral clouds should
increase by a factor 4 from K to J. These ratios are unlikely to be
affected by extinction. The interstellar extinction towards $\rho$~Oph,
excluding intra-nebular extinction, can be estimated from the value of
the colour excess E(B-V) in the maps of \citet{sch98}. Near $\rho$~Oph
but outside any dust clouds extinction is very low, $A_\mathrm{V} <
0.1$, as expected for Gould belt clouds, which raises the J band
intensities by a meagre 3\%.

On the other hand, even at 150~K the C\,{\sc i} continuum is cut off
by its Wien tail above $\sim 5\mu$m, and is $\sim$30~times fainter
than the 8~$\mu$m nebulosity seen in the IRAC~4 image at 8~$\mu$m.  It
is unlikely that the 2MASS images trace a 1000~K C$^+$ plasma because the
solid angle of $\rho$~Oph~W in the K-band image is about 5~arcmin$^2$,
which would give an optically thick continuum of $\sim 10^4$~Jy at
5~GHz \footnote{note, however, that the the faint continuum from the
  diffuse H\,{\sc ii} region about HD147889 is at $\sim$7000~K and
  present in the K-band}.

Is it possible that the near-IR continuum stems from a VSG population
sharing a similar morphology as the PAHs that emit in the IRAC bands?
We note that the circumstellar nebulae around S~1 and SR~3 are also
seen in the 2MASS K$_s$-band images, with JHK colours as for $\rho$~Oph~W
and discrepant from that of scattered light. Sublimation dust
temperatures of $\sim$1000~K are required to explain the S~1 and SR~3
nebulae in terms of large dust grains.  So it is likely that the same
VSGs that would account for the $\rho$~Oph~W near-IR emission would
also be found in S~1 and SR~3. We return to the discussion on spinning
dust in Sec.~\ref{sec:sd}.

\subsubsection{Rovibrational H$_2$}   \label{sec:rovibH2}

An interesting alternative to the VSG continuum is the possibility
that the J, H, and K-band emission in $\rho$~Oph~W stems from H$_2$
rovibrational lines.  \citet[][their Fig.~2]{hab03} present a
(1-0)S(1) H$_2$ image, in which it may be appreciated that
$\rho$~Oph~W is slightly offset by $\sim$10--20~arcsec to the S-E
compared to the ISOCAM PAH emission in the LW2 filter
(5--8.5~$\mu$m). There is also a hint of a shift in
Fig.~\ref{fig:2MASS} between 2MASS K$_s$-band and IRAC~8$\mu$m, which
may be explained if 2MASS~K$_s$ is mostly due to rovibrational H$_2$
and IRAC~8$\mu$m to PAHs. Here we use the {\em Spitzer} spectroscopy
to further investigate relationships between H$_2$, PAHs, and 31~GHz.

The IRS data on $\rho$~Oph~W allows estimating the intensity of the
pure rotational lines H$_2$ S(1) and S(2), in the same manner as used
in Sec.~\ref{sec:sd} for the PAH~11.3~$\mu$m
band. Table~\ref{table:h2} lists the H$_2$ line fluxes in the spectra
we analysed. They are consistent with Table~1 of \citet{hab03}.

\begin{table}
\centering
\caption{{\em Spitzer}~IRS H$_2$ line intensities in W~m$^{-2}$~sr$^{-1}$.}
\label{table:h2}
\begin{tabular}{lrrrr}
\hline 
          & $ \rho$~Oph~W  & SR~3      & S~1              & S~1 off  \\
H$_2$(0-0)S(1) &  3.66(-7)$^a$        &  1.55(-7) &                  &                 \\ 
H$_2$(0-0)S(2) &  2.94(-7)        &  1.86(-7) &   $<$ 1.02(-8)   &    $<$
2.75(-9)  \\  \hline
\end{tabular}
\begin{flushleft}
$^a$  The power of ten exponent is indicated in  parentheses.
\end{flushleft}
\end{table}

A tentative correlation between the H$_2$ fluxes and 31~GHz
intensities can be inferred from a comparison between
Table~\ref{table:h2} and Table~\ref{table:pahratios}. Note that in the
specific case of SR~3, the 31~GHz intensity towards SR~3 should be
$1.4\pm0.2(-1)$, as measured in the 31~GHz image, and not as
tabulated in Table~\ref{table:pahratios} (where we are testing for
31~GHz counterparts to the mid-IR compact source around SR~3).  H$_2$
is weak or undetected in the positions of low 31~GHz intensities
(i.e. S~1 and S~1~off), despite the coincidence with the peak mid- and
far-IR intensities, while H$_2$ is bright at the 31~GHz peak (i.e. on
$\rho$~Oph~W) and in SR~3.  The H$_2$ lines are undetected towards
S~1. The values reported in Table~\ref{table:h2} for S~1 and S~1off
are upper limits derived from formally extracting the H2 line flux,
despite the absence of visible lines.

%                & $ \rho$~Oph~W             & SR~3         & S~1              & S~1 off  \\
%H$_2${\small (0-0)S(2) } &  2.9(-7)       &  1.9(-7)       &   $<$ 1.0(-8)    &    $<$2.7(-9)  \\  
%$I_{31\mathrm{GHz}}$ &  2.2$\pm$0.2(-1) &  1.4$\pm$0.2(-1) &   $<$ 2.4(-2)      &   $<$1.8(-3) \\ \hline 

%A comparison between Table~\ref{table:h2} and
%Table~\ref{table:pahratios} reveals the remarkable fact that H$_2$ is
%weak or undetected in the positions of low 31~GHz intensities
%(i.e. S~1 and SR~3), and is bright at the 31~GHz peak (i.e. on
%$\rho$~Oph~W).
%
In the molecular layers of PDRs the C-ionising UV radiation also
excites the electronic states of H$_2$, which then decay in a
rovibrational cascade leading to the observed near-IR fluorescent
lines. The conditions in $\rho$~Oph~W are very similar to those found
in NGC~2023. \citet{bla87} reproduce the near-IR H$_2$ line system of
NGC~2023 with an incident UV specific intensity field of $I_{UV} =
1240$ in units of the 1000\AA~ UV background at the solar
neighbourhood. For $\rho$~Oph~W we find that $I_{UV} = 545$, if
HD147889 is on the main-sequence, or higher if it is still
contracting. The more recent models of \citet{hab03} demonstrate that
the near-IR H$_2$ lines in $\rho$~Oph~W are fluorescent.

Therefore in the C\,{\sc i} model we expect that the 31~GHz
continuum be coincident with fluorescent H$_2$. The 2MASS-CBI match in
$\rho$~Oph~W could then be explained if the 2MASS filters are
dominated by H$_2$ line emission.

%\citet{hab03} study the H$_2$ spectrum in $\rho$~Oph~W. The IRS fluxes
%are consistent with those observed by ISOCAM.

Given the H$_2$(1-0)S(1) intensity map of \citet{hab03}, we estimate
that ro-vibrational H$_2$ accounts for all of the diffuse 2MASS~K$_s$ flux
seen in $\rho$~Oph~W.  H$_2$(1-0)S(1), on its own, accounts for ~20\%
of the 2MASS K$_s$-band specific intensities. The exact contribution is
difficult to estimate because the diffuse nebulosity in 2MASS is at a
very low level. The noise in 2MASS~K$_s$ is 0.35~MJy~sr$^{-1}$, while the
peak intensities above background in $\rho$~Oph~W are
$\sim$~1MJy~sr$^{-1}$. The H$_2$(1-0)S(1) line, if diluted in the
K$_s$ filter, corresponds to intensities of 0.2~MJy~sr$^{-1}$
\citep[using the H$_2$(1-0)S(1) image in][their
  Fig.~2]{hab03}. Table~3 from \citet[][]{bla87} allows to estimate
that H$_2$(1-0)S(1) represents 21.8\% of the total H$_2$ contribution
in the 2MASS K$_s$ filter.  Therefore H$_2$ ro-vibrational emission
can account for the entire 2MASS K$_s$ band flux.

%H$_2$(1-0)Q(1)~2.406~$\mu$m, at the edge of the K-band, is about as
%bright as S(1), and could contribute equally to the 2MASS
%intensities. **********DOES Q(1) FALL INSIDE THE 2MASS FILTER?
%************
%

%\citet{hab03} fit the pure rotational H$_2$ lines of $\rho$~Oph~W with
%collisional excitation only, in LTE.  They infer a rather high
%temperature, of 300K. It may be that UV pumping also plays a role in
%populating the rotational levels of H$_2$, as suggested by the lack of
%mid-IR H$_2$ lines near S~1.  \citet{hab03} treated HD147889 as a main
%sequence star, with insufficient UV radiation to excite H$_2$ in
%$\rho$~Oph~W.
%

%perldl> $l = 1.25; $Av = 0.1; $EJ_K = $Av/ 5.82; $A_l = $EJ_K * 2.4 *
%($l**(-1.75));
%perldl> p 10**(+$A_l/2.5)
%1.02603541738835
%

%
%\subsection{spectral index}
%
%The spinning dust model requires a 31~GHz spectral index of
%$\alpha_{31} = 1.13, 1.80$, for the 'DRK','RN' cases. For the C\,{\sc
%  i} continuum it would be $\alpha_{31} = 1.36 - $, depending on the
%opacity profile.
%
%steep index
%
%The ff model has the advantage of requiring 

%\section{Conclusion}  \label{sec:conc}

%In the cold plasma model the absence of a circumstellar peak at 31~GHz
%about S~1 is explained by the lack of C-ionising radiation. By
%contrast, spinning dust should be very bright near S~1. 

\subsection{Both C\,{\sc i} and spinning dust?}  \label{sec:discconc}

This work focusses on the 31~GHz continuum as seen by the CBI. Yet
inspection of the {\em WMAP}~Ka map reveals that the entire $\rho$~Oph
cloud is outlined at 33~GHz, albeit at $\sim 1/3$ the intensities in
$\rho$~Oph~W. In particular the eastern filamentary extensions of
$\rho$~Oph, LDN~1729 and LDN~1712, at 0.03~MJy~sr$^{-1}$ and
0.016~MJy~sr$^{-1}$, are not detected in the 2.3~GHz map of
\citet[][in which no region of $\rho$~Oph has a
  counterpart]{baa80}. The diffuse emission surrounding our 45~arcmin
centred on $\rho$~Oph~W, also appears to have a positive spectral
index between 2.3~GHz and 33~GHz $\alpha_{2.3}^{33}$.  If we correct
for a diffuse background at 0.03~MJy~sr$^{-1}$, the WMAP~33~GHz flux
density inside the CBI PB aperture drops from 14~Jy to $\sim$9~Jy.

C\,{\sc i} can better accomodate a reduced 33~GHz flux density, since
the 2.3~GHz -- 33~GHz rise is shallower ($\alpha_{2.3}^{33}$ drops
from 0.69 to 0.52). But what is the origin of the background? The
first possibility that springs to mind is spinning dust, which does
not require an ionisation source, and scales linearly with density.
We could also envision an ensemble of C\,{\sc i}-emitting dense clumps
illuminated by the general interstellar UV field. In this case the
specific intensity in the optically thin regime at 33~GHz can be
written
\begin{equation}
I_\nu = f L n_{\mathrm{C}^+} n_e B_\nu \kappa^1_\nu, 
\end{equation}
where L is the depth of the $\rho$~Oph cloud, $f$ is the filling
factor, and $\kappa^1_\nu$ is as in Eq.~\ref{eq:CIIstromgren}.  An
upper limit to the product $n_{\mathrm{C}^+} n_e$ can be obtained by
ionisation balance in ionisation-bounded clumps
\begin{equation}
A l  n_{\mathrm{C}^+} n_e \alpha <  s_{\mathrm{C}} A, 
\end{equation}
for clumps with depth $l$ and area $A$, where 
\begin{equation}
s_{\mathrm{C}}  = \int_{\lambda_\mathrm{C}}^{\lambda_\mathrm{H}}
d\lambda ~\lambda \,  \pi J_\lambda / (h c),
\end{equation}
is the flux of C-ionising photons per unit area in the interstellar
radiation field $J_\lambda$.  If $ f L \approx l$, the depth of a
single clump, then we have an upper limit to any C\,{\sc i} continuum
diffuse background intensity
\begin{equation}
I^u_\nu = B_\nu \kappa^1_\nu s_\mathrm{C} / \alpha.
\end{equation}
The interstellar radiation field of \citet{mat83} gives $I_\nu^u =
0.4~$MJy~sr$^{-1}$, which is satisfyingly a factor of four above the
observed values in $\rho$~Oph. As in the case of $\rho$~Oph~W
(Sec.~\ref{sec:CI}), we require beam-diluted optically thick clumps
covering a solid angle of $\sim \Omega_\mathrm{beam} T_\mathrm{rms} /
T_e$. Thus to explain the absence of $\rho$~Oph in \citet{baa80} with
a noise of $\sim$30~mK and a beam of 20~arcmin FWHM, the projected
linear size of the clump ensemble should be $<1~$arcmin , if $T_e \sim
100~$K. In $\rho$~Oph~W such narrow dimmensions could perhaps be
interpreted as a very narrow C\,{\sc ii} shell around HD~147889. But
for the diffuse emission such a geometry seems very contrived.

\section{Conclusion}  \label{sec:conc}

We have found that the well-studied and nearby molecular cloud
$\rho$~Oph is surprisingly bright at 31~GHz, or $\sim$1~cm
wavelengths. The most conspicuous feature revealed by the CBI data is
the $\rho$~Oph~W PDR.

%indicate a mass of 922~M$_\odot$ for the $\rho$~Oph cloud

Comparison with {\em~WMAP} images shows that the cm-emission is not
the Rayleigh-Jeans tail of the sub-mm emitting dust. Bulk dust
properties are inferred from new {\em ISO}-LWS parallel mode
data. Archival {\em Spitzer} data allows quantifying PAH
emission. None of the comparison images match the CBI data, except for
IRAC~8~$\mu$m and 2MASS~K$_s$, which closely follow the 31~GHz ridge
along $\rho$~Oph~W. {\em Spitzer}~IRS spectroscopy in four apertures
hint at a possible correlation between 31~GHz intensity and the H$_2$
pure-rotational lines: the H$_2$ lines are detected only where 31~GHz
intensities are bright (i.e. towards SR~3 and $\rho$~Oph~W, and not in
the vicinity of S~1).

We considered several interpretations for the 31~GHz emision,
requiring either an additional radio continuum component such as
magnetic or spinning dust, or unexpected physical conditions, such as
in a very dense and cold plasma. We find that both spinning dust and
C\,{\sc i} continuum can explain the data:

\begin{itemize}

\item Magnetic dust is discarded on morphological grounds. The
  expected 31~GHz intensity from a magnetic enhancement of the grain
  opacity is at odds with the CBI data. The polarization levels
  required by magnetic dust are not observed.

%We considered several interpretations, requiring either an additional
%radio continuum component  such as spinning dust, or unexpected
%physical conditions, such as an ensemble of dense ridges giving rise
%to cold plasma emission. 

\item Spinning dust can account for the radio spectrum. But the
  predicted levels in S~1 are in excess by a factor $>$40 at
  3~$\sigma$.  We take the intensities in PAH~11.3~$\mu$m as a proxy
  for the mid-IR VSG emission, which is approximately proportional to
  both the VSG column and local UV intensity $U$.  Taking into account
  the range of spinning dust emissivities in all possible
  environments, the uncertainties in our 31~GHz map can marginally
  reconcile spinning dust with the data (i.e. at 3~$\sigma$
  variations).

%Allowing for VSG depletion also decreases the mid-IR emission.

% For spinning dust the product of $U$ times $R$, the ratio of 31~GHz
% to PAH~11.3~$\mu$m intensities, should not vary by a factor larger
% than the ratio of spinning dust emissivities in different
% environments, which is $\sim$10 \citep{dl98b}. A comparison of PAH
% intensities measured by {\em Spitzer}~IRS in S~1, SR~3 and
% $\rho$~Oph~W, shows that $U \times R$ varies by a factor of up to 27,
% which given the uncertainties could be as low as 4.
%

\item Alternatively a cold plasma such as that found in C\,{\sc ii}
  regions could explain the 31~GHz emission. The star HD147889, a
  binary pre-main-sequence star with spectral types B2IV, B3IV, emits
  sufficient C-ionising UV radiation to interpret $\rho$~Oph~W as a
  C\,{\sc ii} region.  But the absence of detectable $<10~$GHz signal
  requires optically thick emission, and hence high H-nucleus
  densities ($n_\mathrm{H} \sim 10^7$~cm$^{-3}$, almost a factor of 10
  higher than inferred in the literature), implying a hitherto
  unobserved molecular phase of the ISM. In $\rho$~Oph this molecular
  phase takes the form of an ensemble of dense clumps or sheets at
  temperatures of order 50~K or less.  The cold plasma interpretation
  explains the 2MASS - CBI correlation through an important
  fluorescent H$_2$ contribution to the 2MASS bands.

\end{itemize}

%Need accurate 5GHz data to conclude, as well as K-band spectroscpy and
%polarimetry. 

\section*{Acknowledgments}

We thank the referee for a thorough reading and useful
comments. S.C. acknowledges support from FONDECYT grant 1060827, and
from the Chilean Center for Astrophysics FONDAP 15010003.  We
gratefully acknowledge the generous support of Maxine and Ronald
Linde, Cecil and Sally Drinkward, Barbara and Stanely Rawn, Jr., Fred
Kavli, and Rochus Vogt.  This work was supported by the National
Science Foundation under grants AST 00-98734 and AST 02-06416.  RP
acknowledges the support of a Spitzer Cycle-5 archival proposal grant
(PAC.PALADINI - 1 - JPL.000094). This publication makes use of data
products from 1- the Two Micron All Sky Survey, which is a joint
project of the University of Massachusetts and the Infrared Processing
and Analysis Center, funded by the National Aeronautics and Space
Administration and the National Science Foundation, and 2- the
Southern H-Alpha Sky Survey Atlas (SHASSA), which is supported by the
National Science Foundation. The Mopra radio telescope is part of the
Australia Telescope which is funded by the Commonwealth of Australia
for operations as a National Facility operated by CSIRO. This research
also made use of Montage, funded by the National Aeronautics and Space
Administration's Earth Science Technology Office, Computational
Technnologies Project, under Cooperative Agreement Number NCC5-626
between NASA and the California Institute of Technology. The code is
maintained by the NASA/IPAC Infrared Science Archive. This research
has made use of the SIMBAD database, operated at CDS, Strasbourg,
France.

\appendix

\section{Image Reconstruction} \label{sec:imagereconstruction}

\begin{figure}
\begin{center}
\includegraphics[width=0.7\columnwidth,height=!]{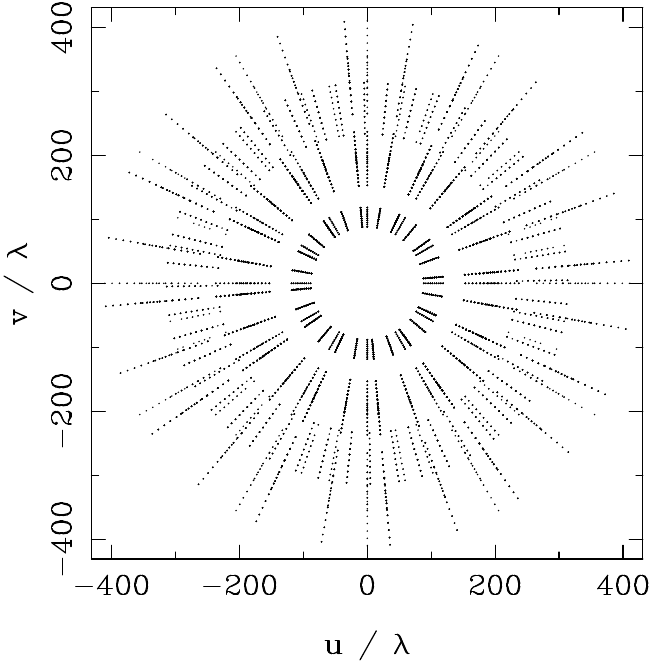}
% uvcov.pl
\end{center}
\caption{\label{fig:uvcov}
$(u,v)$ coverage of the CBI in the compact configuration used
  for the observations of $\rho$~Oph~W.   }
\end{figure}

Image reconstruction from interferometer data is an instance of the
inverse problem.  The compact configuration of the CBI interferometer
results in the $(u,v)$ coverage shown in Fig.~\ref{fig:uvcov}. Missing
spacings and noise impose non-linear deconvolution methods.  With a
uniform grid the number of free parameters (here $256^2$) can exceed
the number of independent data points, so that the fit is degenerate.
In this case we take as a measure of the quality of fit $\chi^2/f$,
where $f$ is the number of independent data points.  $\chi^2/f$ should
be close to 1 for a satisfactory model.  The first few conjugate
gradient iterations in the least-squares algorithm fit the bulk of the
signal, but inevitably converge on noisy models, with $\chi^2/f$
values much less than 1.  MEM regularization prevents the models from
fitting the noise.  The model shown in Fig.~\ref{fig:CBI0} has
$\chi^2/f = 1.20$, which is somewhat high, but reconcilable with 1.0
through a 10\% increase in the noise. The noise in each visibility
point is calculated as the root-mean-square (rms) scatter of $\sim$120
samples, and does not include residual errors in calibration, which
may perhaps amount to 10\%.

A reconstruction starting from a blank image as initial condition,
rather than the image prior $M_i$, converges on essentially the same
model as Fig.~\ref{fig:CBI0}, but with a higher $\chi^2/f = 1.4$, and
thus larger residuals.  The prior we chose accounts for some of the
extended emission seen by {\em WMAP}, but keeps the IRAC resolution by
replacing the emission within the CBI primary beam with the
IRAC~8~$\mu$m image, suitably scaled and processed so as to resemble
the blank-prior MEM models (i.e. we patched-out stellar features,
namely S~1 and SR~3, see below). This patched image was then fed as
initial condition and prior into a sky-plane MEM deconvolution of the
{\em WMAP}~Ka image, producing the prior image used for the CBI
reconstructions in Fig.~\ref{fig:CBI0}, with $\chi^2/f = 1.2$.

\begin{figure}
\begin{center}
\includegraphics[width=\columnwidth,height=!]{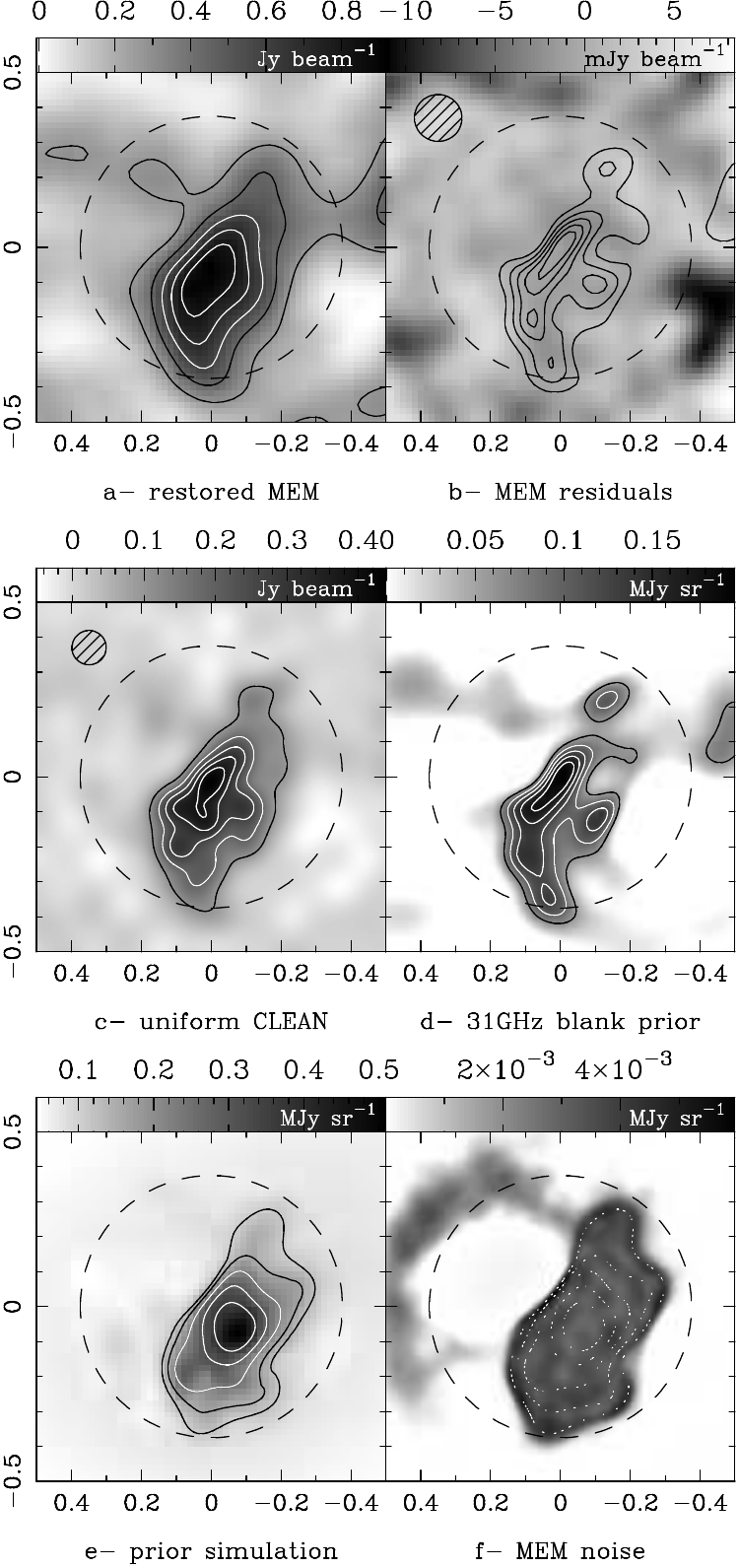}
% image_roph_CBI_res_all.pl
\end{center}
\caption{\label{fig:CBI} MEM restored image and comparison
  reconstructions.  The axes and dashed circle follow from
  Fig.~\ref{fig:CBI0}.  {\bf a-} restored image, with contour levels
  at 0.012, 0.286, 0.446, 0.581, 0.702 and 0.813 Jy~beam$^{-1}$, and
  peak at 0.916~Jy~beam$^{-1}$. {\bf b-} MEM contours overlaid on the
  residual image. The natural-weight synthetic beam is indicated by
  the hatched ellipse.  {\bf c-} {\sc difmap clean} reconstruction of
  the CBI data, in uniform weights. Contour levels are 0.098, 0.185,
  0.257, 0.323 and 0.382 Jy~beam$^{-1}$.  {\bf d-} A CBI MEM model
  with a blank prior is shown in grey scale, with contours at 0.06,
  0.09, 0.12, 0.15 and 0.17~MJy~sr$^{-1}$.  {\bf e-} average MEM model
  of CBI simulations on the prior image used to produce our best
  model.  Contours are at 0.098, 0.185, 0.257, 0.323,
  0.382~MJy~sr$^{-1}$, overlaid on a grey scale of the input
  image. {\bf f-} averaged MEM contours overlaid on the
  root-mean-square (rms) dispersion of the MEM models.}
\end{figure}

The CBI observations and the quality of the MEM model can be examined
through the restored image and the residuals shown on
Fig.~\ref{fig:CBI}a, ~\ref{fig:CBI}b.  The residual image is the dirty
map \citep[produced with the DIFMAP package using natural
  weights,][]{she97}, of the residual visibilities (the difference
between the observed and model visibilities). The restored image is
the sum of the residual image, after division by the CBI primary beam,
and the MEM model convolved with an elliptical Gaussian fit to the
natural-weight synthetic beam (8.17$\times $7.99~arcmin$^2$). In
natural weights the theoretical rms noise of the dirty map is
3.4~mJy~beam$^{-1}$. The dynamic range of the restored image shown in
Fig.~\ref{fig:CBI} is of order 100.

We examined the statistical properties of the MEM reconstructions with
the simulations described in Appendix~B of \citet{cas06}. These
simulations are relevant to assess the effects of the CBI beam on the
morphological analysis of Sec.~\ref{sec:morph}, and to estimate the
dynamic range of the MEM reconstructions.  The CBI MEM prior is a
suitable template.  We ran the same MEM algorithm as applied to the
CBI visibilities on a simulation of CBI observations on the prior used
for the reconstruction of Fig.~\ref{fig:CBI0}.  Fig.~\ref{fig:CBI}
compares this prior image with the average and dispersion of the MEM
models run on 90 realisations of complex visibility noise.  Note that
in this case we used a blank prior with $\lambda = 20$ and $M_i =
0.3~$Jy~sr$^{-1}$.  The 31~GHz model image obtained with this MEM
reconstruction is shown in Fig.~\ref{fig:CBI}d. It is very similar to
Fig.~\ref{fig:CBI0}. We take this similarity as a proof of robustness
of the MEM reconstructions, and also as ground to extrapolate the
dynamic range of the simulations, of $\sim$20, to the CBI images.

%IRAC~8~$\mu$m image of $\rho$~Oph, after scaling
%the IRAC visibilities to the CBI values by linear regression.

%%%[io:/io2/mem_output/ROPHW_2007/64/CBI_l20_ftol1e-5/mem/]%

%The dirty map
%of MEM residuals, for any particular realisation of noise, also
%includes significant extended negatives, as in Fig.~\ref{fig:CBI}.
% 

%, which is why we show here for comparison the CBI model without
%the use of a prior

%average MEM simulations:
%~simon/mem_working64/perl/avnruns.pl
%[io:~/mem_working64/perl/]% ls -rtl *fits
%-rw-r--r-- 1 simon users 532800 2008-04-08 15:35 MEM_ROPHW_IRAC_simul_av.fits
%-rw-r--r-- 1 simon users 529920 2008-04-08 15:35 MEM_ROPHW_IRAC_simul_rms.fits
%[io:~/mem_working64/perl/]% 
%perl toJy.pl ./MEM_ROPHW_IRAC_simul_rms.fits   ./MEM_ROPHW_IRAC_simul_rms_Jy.fits 
%perl toJy.pl ./MEM_ROPHW_IRAC_simul_av.fits   ./MEM_ROPHW_IRAC_simul_av_Jy.fits 
%

%\begin{figure}
%\begin{center}
%\includegraphics[width=\columnwidth,height=!]{roph_IRAC_simul.pdf}
%% perl image_roph_IRAC_simul.pl
%\end{center}
%\caption{\label{fig:simul} }
%\end{figure}
%

%#  Tue-Wed 03-04 August 2004:  NO & JN
%#  Wed-Thu 04-05 August 2004:  NO & JC
%#  --------------------------------------------------------------------------------
%#  Mon-Tue 04-05 October 2004
%#  Sun-Mon 26-27 September 2004:  RR & NO.
%

\label{lastpage}

\end{document}